\documentclass[a4paper,11pt]{article}
\pdfoutput=1
\usepackage{jheppub}

\usepackage{amsmath}
\usepackage{amssymb}
\usepackage{bm}
\usepackage{xspace}
\usepackage{afterpage}
\usepackage{listings}
\usepackage{subcaption}

\usepackage[T1]{fontenc}

\usepackage{titlesec}
\titleformat*{\subsection}{\bfseries\boldmath}
\titleformat*{\subsubsection}{\bfseries\boldmath}

\usepackage{esvect} 

\usepackage{enumitem}

\usepackage{hyperref}

\usepackage[bottom]{footmisc}

\title{\boldmath Alternative angular variables for suppression of QCD
  multijet events in new physics searches with missing transverse
  momentum at the LHC}

\author[a]{Tai Sakuma,}
\author[a]{Henning Flaecher,}
\author[a,b]{and Dominic Smith}
\affiliation[a]{University of Bristol, Bristol, UK}
\affiliation[b]{Vrije Universiteit Brussel, Brussel, Belgium}

\emailAdd{tai.sakuma@bristol.ac.uk}

\abstract{We introduce three alternative angular variables---denoted
  by $\tilde{\omega}_\text{min}$, $\hat{\omega}_\text{min}$, and
  $\chi_\text{min}$---for QCD multijet event suppression in
  supersymmetry searches in events with large missing transverse
  momentum in proton-proton collisions at the LHC at CERN. In searches
  in all-hadronic final states in the CMS and ATLAS experiments, the
  angle $\Delta\varphi_i$, the azimuthal angle between a jet and the
  missing transverse momentum, is widely used to reduce QCD multijet
  background events with large missing transverse momentum, which is
  primarily caused by a jet momentum mismeasurement or neutrinos in
  hadron decays---the missing transverse momentum is aligned with a
  jet. A related angular variable---denoted by
  $\Delta\varphi^*_\text{min}$, the minimum of the azimuthal angles
  between a jet and the transverse momentum imbalance of the
  \textit{other} jets in the event---is used instead in a series of
  searches in all-hadronic final states in CMS to suppress QCD
  multijet background events to a negligible level. In this paper,
  before introducing the alternative variables, we review the variable
  $\Delta\varphi^*_\text{min}$ in detail and identify room for
  improvement, in particular, to maintain good acceptances for signal
  models with high jet multiplicity final states. Furthermore, we
  demonstrate with simulated event samples that
  $\hat{\omega}_\text{min}$ and $\chi_\text{min}$ considerably
  outperform $\Delta\varphi^*_\text{min}$ and $\Delta\varphi_i$ in
  rejecting QCD multijet background events and that
  $\hat{\omega}_\text{min}$ and $\tilde{\omega}_\text{min}$ are also
  useful for reducing the total standard model background events.}

\DeclareMathOperator{\arccot}{arccot}

\begin{document} 

\graphicspath{{images/}}
\DeclareGraphicsExtensions{.pdf,.png}

\newcommand{\pTsca}{\ensuremath{p^{}_{\text{T}}}\xspace}
\newcommand{\pTvec}{\ensuremath{{\vec p}^{}_{\text{T}}}\xspace}
\newcommand{\pTisca}{\ensuremath{p_{\text{T}i}}\xspace}
\newcommand{\pTivec}{\ensuremath{{\vec p}^{}_{\text{T}i}}\xspace}
\newcommand{\pTjvec}{\ensuremath{{\vec p}^{}_{\text{T}j}}\xspace}
\newcommand{\pTscaJet}{\ensuremath{p^{\text{jet}}_{\text{T}}}\xspace}
\newcommand{\pTvecJet}{\ensuremath{{\vec p}^{\kern2pt\text{jet}}_{\text{T}}}\xspace}

\newcommand{\pTivecO}{\ensuremath{{\vec p}^{\kern1pt\,0}_{\text{T}i}}\xspace}

\newcommand{\pToneSca}{\ensuremath{p^{}_{\text{T}1}}\xspace}
\newcommand{\pToneVec}{\ensuremath{{\vec p}^{}_{\text{T}1}}\xspace}
\newcommand{\pTtwoSca}{\ensuremath{p^{}_{\text{T}2}}\xspace}
\newcommand{\pTtwoVec}{\ensuremath{{\vec p}^{}_{\text{T}2}}\xspace}
\newcommand{\pTthreeSca}{\ensuremath{p^{}_{\text{T}3}}\xspace}
\newcommand{\pTthreeVec}{\ensuremath{{\vec p}^{}_{\text{T}3}}\xspace}

\newcommand{\pToneTrueSca}{\ensuremath{p^{\text{true}}_{\text{T}1}}\xspace}
\newcommand{\pToneTrueVec}{\ensuremath{{\vec p}^{\;\text{true}}_{\text{T}1}}\xspace}
\newcommand{\pToneOverSca}{\ensuremath{p^{\text{over}}_{\text{T}1}}\xspace}
\newcommand{\pToneOverVec}{\ensuremath{{\vec p}^{\;\text{over}}_{\text{T}1}}\xspace}

\newcommand{\HT}{\ensuremath{H_{\text{T}}}\xspace}

\newcommand{\njet}{\ensuremath{n_{\text{jet}}}\xspace}

\newcommand{\MpTsca}{\ensuremath{p_{\text{T}}^{\text{miss}}}\xspace}
\newcommand{\MpTvec}{\ensuremath{{\vec p}_{\text{T}}^{\kern2pt\text{miss}}}\xspace}

\newcommand{\METsca}{\ensuremath{E_{\text{T}}^{\text{miss}}}\xspace}
\newcommand{\METvec}{\ensuremath{{\vec E}_{\text{T}}^{\text{miss}}}\xspace}

\newcommand{\MHTsca}{\ensuremath{H_{\text{T}}^{\text{miss}}}\xspace}
\newcommand{\MHTvec}{\ensuremath{{\vec H}_{\text{T}}^{\text{miss}}}\xspace}

\newcommand{\MHTvecO}{\ensuremath{{\vec H}_{\text{T}}^{\text{miss}\,0}}\xspace}

\newcommand{\bMHTsca}{\ensuremath{H_{\text{T}}^{\text{miss}*}}\xspace}
\newcommand{\bMHTvec}{\ensuremath{{\vec H}_{\text{T}}^{\text{miss}*}}\xspace}

\newcommand{\bMHTisca}{\ensuremath{H_{\text{T}i}^{\text{miss}*}}\xspace}
\newcommand{\bMHTivec}{\ensuremath{{\vec H}_{\text{T}i}^{\text{miss}*}}\xspace}

\newcommand{\bMHToneSca}{\ensuremath{H_{\text{T}i}^{\text{miss}*}}\xspace}
\newcommand{\bMHToneVec}{\ensuremath{{\vec H}_{\text{T}1}^{\text{miss}*}}\xspace}

\newcommand{\mmMHTsca}{\ensuremath{\widetilde{H}_{\text{Tmin}}^{\text{miss}}}\xspace}
\newcommand{\mmMHTvec}{\ensuremath{{\vec{\widetilde{H}}}{}_{\text{Tmin}}^{\text{miss}}}\xspace}

\newcommand{\Dphi}{\ensuremath{\Delta\varphi}\xspace}
\newcommand{\bDphi}{\ensuremath{\Delta\varphi^*}\xspace}
\newcommand{\DphiTilde}{\ensuremath{\Delta\tilde{\varphi}}\xspace}
\newcommand{\DphiHat}{\ensuremath{\Delta\hat{\varphi}}\xspace}
\newcommand{\bDphiTilde}{\ensuremath{\Delta\tilde{\varphi}^*}\xspace}
\newcommand{\omegaTilde}{\ensuremath{\tilde{\omega}\xspace}}
\newcommand{\omegaHat}{\ensuremath{\hat{\omega}\xspace}}

\newcommand{\Dphii}{\ensuremath{\Dphi_i}\xspace}
\newcommand{\DphiTildei}{\ensuremath{\DphiTilde_i}\xspace}
\newcommand{\DphiHati}{\ensuremath{\DphiHat_i}\xspace}
\newcommand{\bDphii}{\ensuremath{\bDphi_i}\xspace}
\newcommand{\bDphiTildei}{\ensuremath{\bDphiTilde_i}\xspace}
\newcommand{\omegai}{\ensuremath{\omega_i}\xspace}
\newcommand{\omegaTildei}{\ensuremath{\omegaTilde_i}\xspace}
\newcommand{\omegaHati}{\ensuremath{\omegaHat_i}\xspace}
\newcommand{\chii}{\ensuremath{\chi_i}\xspace}

\newcommand{\Dphione}{\ensuremath{\Dphi_1}\xspace}
\newcommand{\bDphione}{\ensuremath{\bDphi_1}\xspace}

\newcommand{\maxf}{\ensuremath{f_\text{max}}\xspace}
\newcommand{\minbDphi}{\ensuremath{\bDphi_\text{min}}\xspace}
\newcommand{\minDphi}{\ensuremath{\Dphi_\text{min}}\xspace}
\newcommand{\minDphiFour}{\ensuremath{\Dphi_\text{min4}}\xspace}
\newcommand{\minOmega}{\ensuremath{\omega_\text{min}}\xspace}
\newcommand{\minOmegaTilde}{\ensuremath{\tilde{\omega}_\text{min}}\xspace}
\newcommand{\minOmegaHat}{\ensuremath{\hat{\omega}_\text{min}}\xspace}
\newcommand{\minChi}{\ensuremath{\chi_\text{min}}\xspace}

\newcommand{\fii}{\ensuremath{f_i}\xspace}
\newcommand{\gii}{\ensuremath{g_i}\xspace}
\newcommand{\kii}{\ensuremath{k_i}\xspace}
\newcommand{\hii}{\ensuremath{h_i}\xspace}

\newcommand{\fione}{\ensuremath{f_1}\xspace}

\newcommand{\alphat}{\ensuremath{\alpha_{\text{T}}}\xspace}
\newcommand{\MTtwo}{\ensuremath{M_\text{T2}}\xspace}

\newcommand{\OAvec}{\ensuremath{\protect\vv{\text{OA}}}\xspace}
\newcommand{\OBvec}{\ensuremath{\protect\vv{\text{OB}}}\xspace}
\newcommand{\OCvec}{\ensuremath{\protect\vv{\text{OC}}}\xspace}
\newcommand{\CBvec}{\ensuremath{\protect\vv{\text{CB}}}\xspace}
\newcommand{\BCvec}{\ensuremath{\protect\vv{\text{BC}}}\xspace}
\newcommand{\ODvec}{\ensuremath{\protect\vv{\text{OD}}}\xspace}
\newcommand{\DBvec}{\ensuremath{\protect\vv{\text{DB}}}\xspace}

\newcommand{\MADGRAPH} {\textsc{MadGraph}\xspace}
\newcommand{\MADSPIN} {\textsc{MadSpin}\xspace}
\newcommand{\MCATNLO} {\textsc{mc@nlo}\xspace}
\newcommand{\MGvATNLO}{\MADGRAPH{}5\_a\MCATNLO{}\xspace}
\newcommand{\PYTHIA} {{\textsc{pythia}}\xspace}
\newcommand{\FASTJET} {{\textsc{FastJet}}\xspace}
\newcommand{\DELPHES} {{\textsc{delphes}}\xspace}

\newcommand{\ttbar}{\ensuremath{\mathrm{t}\overline{\mathrm{t}}}\xspace} 
\newcommand{\ttjets}{{{\ttbar}+jets}\xspace}
\newcommand{\PW}{\ensuremath{\mathrm{W}}\xspace}
\newcommand{\cPZ}{\ensuremath{\mathrm{Z}}} 
\newcommand{\cPgn}{\ensuremath{\nu}} 
\newcommand{\cPagn}{\ensuremath{\overline{\nu}}} 
\newcommand{\znnjets}{{\ensuremath{\cPZ(\to\cPgn \cPagn )}+jets}\xspace}
\newcommand{\wjets}{{{\PW}+jets}\xspace}

\newcommand{\PSg}{\ensuremath{\mathrm{\widetilde{g}}}} 
\newcommand{\PSq}{\ensuremath{{\widetilde{q}}}} 
\newcommand{\PSqt}{\ensuremath{{\widetilde{t}}}} 
\newcommand{\PSqb}{\ensuremath{{\widetilde{b}}}} 
\newcommand{\PSGczDo}{\ensuremath{{\widetilde{\chi}^0_\mathrm{1}}}} 

\maketitle
\flushbottom

\section{\boldmath Introduction}
\label{sec:intro}

Supersymmetry (SUSY) searches in proton-proton collisions are ongoing
at the CMS and ATLAS experiments at the LHC; recent results of
searches in all-hadronic final states can be found, for example, in
Refs.~\cite{Sirunyan:2017cwe, Sirunyan:2017kqq, Sirunyan:2017wif,
  Sirunyan:2017kiw, Aaboud:2017hdf, Aaboud:2017wqg, Aaboud:2017ayj,
  Sirunyan:2017obz, Sirunyan:2017pjw, Aaboud:2017hrg, Aaboud:2017phn,
  Aaboud:2017vwy, Sirunyan:2017bsh, Sirunyan:2018vjp}. In SUSY models
with $R$-parity conservation, SUSY particles are produced in pairs;
then, their cascade decay chains each end with the lightest
supersymmetric particle (LSP), which is typically the neutralino,
\textit{invisible} to the detectors, resulting in missing transverse
momentum (denoted by \METvec and its magnitude by \METsca). For this
reason, SUSY particles are usually searched for in events with large
\METsca. In all-hadronic final states, \METvec can be approximated by
the missing transverse hadronic momentum (denoted by \MHTvec and its
magnitude by \MHTsca), the negative vector sum of the transverse
momenta (\pTvec) of all reconstructed \textit{jets} in the event.
Instead of or in addition to \METvec, \MHTvec is often used in
CMS~\cite{Sirunyan:2017cwe, Sirunyan:2017kqq, Sirunyan:2018vjp}.

In searches in all-hadronic final states, the angle \Dphii---the
azimuthal angle between a jet $i$ and either \METvec or \MHTvec---is
widely used to reduce QCD multijet background events with large
\METsca, which is primarily caused by a jet mismeasurement or by
neutrinos in hadron decays in a jet~\cite{Sirunyan:2017cwe,
  Sirunyan:2017kqq, Sirunyan:2017wif, Sirunyan:2017kiw,
  Aaboud:2017wqg, Aaboud:2017ayj, Sirunyan:2017obz, Sirunyan:2017pjw,
  Aaboud:2017hrg, Aaboud:2017phn, Aaboud:2017vwy, Sirunyan:2017bsh}.
The angle \Dphii is narrow for the jet whose \pTsca is
\textit{underestimated} because of either a mismeasurement or
neutrinos in hadron decays.
In searches using \Dphii, it is common to require the angles \Dphii of
only a few highest \pTsca jets in the event to be wider than
thresholds, and the thresholds are sometimes narrower for jets with
lower \pTsca rankings in the event~\cite{Sirunyan:2017cwe,
  Sirunyan:2017bsh, Sirunyan:2017pjw, Sirunyan:2017wif,
  Sirunyan:2017obz, Aaboud:2017vwy}. For example, the search in
Ref.~\cite{Sirunyan:2017cwe} required $\Dphii > 0.5$ for the two
highest \pTsca jets and $\Dphii > 0.3$ for the 3rd and 4th highest
\pTsca jets and had no requirement for the other jets.

It is uncommon to apply a lower threshold on the minimum of the angles
\Dphii of all jets in the event, in other words, to require all jets
in the event to have \Dphii wider than a certain angle, especially in
high jet multiplicity events. This is because such a requirement would
reduce too much the signal acceptances of SUSY models. In particular,
signal events with high jet multiplicity have high chances to have at
least one jet with narrow \Dphii. Consequently, in common practice, a
QCD multijet event with large \METsca caused by a jet \pTsca
underestimate will not be rejected unless the jet whose \pTsca is
underestimated is among the highest \pTsca jets in the event.

On the other hand, if the large \METsca in a QCD multijet event is due
to a jet \pTsca \textit{overestimate}, the angle \Dphii of the jet
whose \pTsca is overestimated is wide (near $\pi$). However, it is
also uncommon to apply an upper threshold on \Dphii because this would
also reduce signal acceptances too much. For example, jets in signal
events that receive the recoil momenta of the LSPs have often wide
\Dphii. As a result, in common practice, a QCD multijet event with
large \METsca caused by a jet \pTsca overestimate will not be rejected
either.

In Refs.~\cite{Khachatryan:2011tk, Khachatryan:2016pxa,
  Khachatryan:2016dvc, Sirunyan:2018vjp}, CMS used the angular
variable \minbDphi, the minimum of the angles \bDphii of all jets in
the event, where \bDphii denotes the azimuthal angle between a jet $i$
and the negative vector sum of \pTvec of all \textit{other} jets in
the event. The angular variable \minbDphi first appeared in
Refs.~\cite{CMS-PAS-SUS-09-001, CMS-PAS-SUS-10-001} and was used to
identify large \METsca caused by masked regions of the
calorimeter~\cite{Khachatryan:2011tk}.\footnote{In
  Ref.~\cite{CMS-PAS-SUS-09-001}, \minbDphi is referred to as
  ``biased'' $\Delta\phi$. In Refs.~\cite{CMS-PAS-SUS-10-001,
    Khachatryan:2011tk}, \minbDphi is denoted by $\Delta\phi^*$.} In
recent results~\cite{Khachatryan:2016pxa, Khachatryan:2016dvc,
  Sirunyan:2018vjp}, \minbDphi was used, along with another
dimensionless variable $\alpha_\text{T}$~\cite{Randall:2008rw,
  Khachatryan:2011tk}, to suppress QCD multijet background events to a
negligible level; events were required to have \minbDphi above a
threshold, $\minbDphi \ge \gamma_0$ (the \textit{$\mathit{\minbDphi}$
  criterion}), with $\gamma_0 = 0.3$ in
Ref.~\cite{Khachatryan:2016pxa} and 0.5 in
Refs.~\cite{Khachatryan:2016dvc, Sirunyan:2018vjp}.

The \minbDphi criterion can reject QCD multijet background events with
large \METsca caused by a jet \pTsca overestimate or by an
underestimate even when the jet whose \pTsca is underestimated is not
among the highest \pTsca jets in the event. The \minbDphi criterion
can reject QCD multijet background events to a negligible level while
still keeping signal acceptances of SUSY models large enough to carry
out the search. However, the \minbDphi criterion does largely reduce
signal acceptances, in particular of models with high jet
multiplicity.\footnote{See, for example, the additional tables 7-11
  of~\cite{Khachatryan:2016dvc}, which can be found at
  \url{http://cms-results.web.cern.ch/cms-results/public-results/publications/SUS-15-005/index.html}.}

In this paper, after reviewing the angle \bDphii and the \minbDphi
criterion, we introduce three alternative angular variables:
\minOmegaTilde, \minOmegaHat, and \minChi. These variables are
designed to maintain good signal acceptances of SUSY models, in
particular of models with high jet multiplicity in the final state,
while rejecting QCD multijet background events to a negligible level.
Receiver operating characteristic (ROC) curves in simulated event
samples demonstrate that \minOmegaHat and \minChi considerably
outperform \minbDphi and \Dphii in rejecting QCD multijet background
events and that \minOmegaHat and \minOmegaTilde are also useful for
reducing the total standard model background events.

This paper proceeds as follows. Section~2 reviews the angle \bDphii.
In Section~3, we determine the minimized \MHTsca in the variation of
jet \pTsca and show the relation of the minimized \MHTsca to the angle
\bDphii. Section~4 describes simulated event samples. In Section~5, we
show that the minimized \MHTsca for the jet that is mismeasured or has
neutrinos from hadron decays is small in QCD multijet events in the
simulated event samples. Section~6 reviews features of the \minbDphi
criterion and identifies room for improvement. The three alternative
angular variables, \minOmegaTilde, \minOmegaHat, and \minChi, are
introduced in Section~7. Section~8 demonstrates the performance of
these variables in the simulated event samples. A summary is given in
Section~9.

\section{\boldmath Review of angle \texorpdfstring{\bDphii}{Dphii*}}
\label{sec:bdphi}

In this section, after providing the definitions of kinematic
variables, including \bDphii, as used in this paper, we show that
\bDphii can be written as a function of two dimensionless variables
and review the properties of \bDphii.

\subsection{Definitions}

\paragraph{\boldmath Missing transverse hadronic momentum \MHTvec.}
The missing transverse hadronic momentum (\MHTvec) is the negative
vector sum of \pTvec of all jets in the event:
\begin{align}
  \label{eq:mht}
  \MHTvec \equiv -\sum_{i \in \text{jets}}\pTivec.
\end{align}
Its magnitude is denoted by \MHTsca. The two vectors \METvec and
\MHTvec have similar values in all-hadronic final states. In this
paper, we focus on all-hadronic final states and primarily use
\MHTvec. However, much of the discussion in this paper is valid even
if \MHTvec is replaced with \METvec.

\paragraph{\boldmath Angle \Dphii.}
The angle \Dphii is the angle between the \pTvec of jet $i$ and
\MHTvec:
\begin{align}
  \label{eq:dphi}
  \Dphii \equiv \Dphi(\pTivec, \MHTvec).
\end{align}
The angle \Dphii ranges from 0 to $\pi$. This angle is often defined
with \METvec instead of \MHTvec. Both definitions have similar values
in all-hadronic final states. We define \Dphii with \MHTvec as in
Eq.~\eqref{eq:dphi} in this paper.

\bigbreak

\noindent
We let \bMHTivec denote the vector sum of \MHTvec and the jet \pTivec:
\begin{align}
  \label{eq:bMHTivec}
  \bMHTivec
  \equiv \MHTvec + \pTivec
  = -\sum_{j \in \text{jets}}\pTjvec + \pTivec
  = -\sum_{\substack{j \in \text{jets} \\ j \neq i}}\pTjvec.
\end{align}
It is the negative vector sum of \pTvec of all jets but the jet $i$ in
the event. Its magnitude is denoted by \bMHTisca.

\paragraph{\boldmath Angle \bDphii.}
The angle \bDphii is the angle between the jet \pTivec and \bMHTivec:
\begin{align}
  \label{eq:bdphi}
  \bDphii \equiv \Dphi(\pTivec, \bMHTivec)
  = \Dphi(\pTivec, \MHTvec + \pTivec).
\end{align}
The angle \bDphii ranges from 0 to $\pi$. In the \METvec-based
definition, this angle can be defined as the angle between the jet
\pTivec and the vector sum $\METvec + \pTivec$. In this paper we
define \bDphii with \MHTvec as in Eq.~\eqref{eq:bdphi}.

\subsection{Geometric relations}
\label{subsec:geo_bdphi_dphi_fii}

\begin{figure}[!h]
\centering
  \begin{subfigure}[t]{0.48\textwidth}
    \centering
    \includegraphics[scale=0.85]{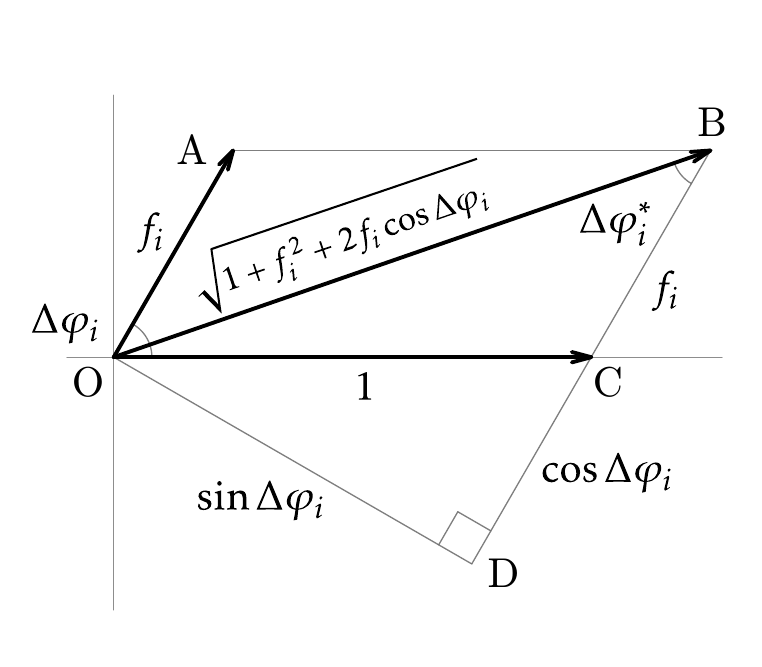}
    \caption{\label{fig:f040_010_015_001} $\Dphii < \pi/2$, a near
      side jet.}
  \end{subfigure}
  ~
  \begin{subfigure}[t]{0.48\textwidth}
    \centering
    \includegraphics[scale=0.85]{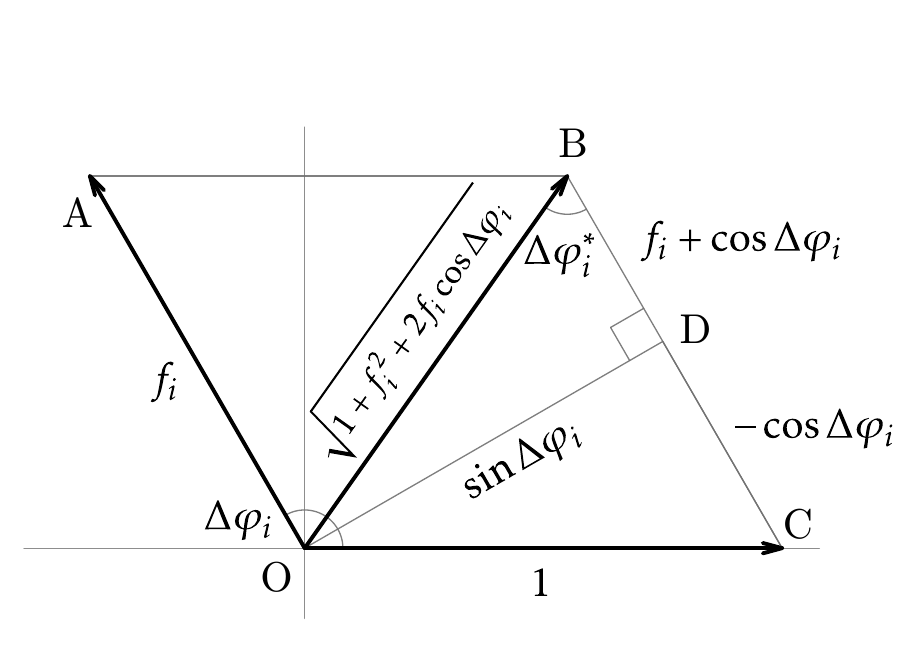}
    \caption{\label{fig:f040_030_015_001} $\Dphii > \pi/2$, an away
      side jet.}
  \end{subfigure}
  \caption{\label{fig:f040_xxx_015_001} The geometric relations among
    \bDphii, \Dphii, and \fii in the normalized \pTsca plane, the
    transverse momentum plane that is rotated and scaled such that
    \MHTvec points horizontally to the right with unit length and that
    is flipped if necessary so that \Dphii lies between 0 and $\pi$.
    The vectors \OAvec, \OBvec, and \OCvec are the jet \pTivec,
    \bMHTivec, and \MHTvec, respectively. The point~D is the
    projection of the point O onto the line BC.}
\end{figure}

Since \bDphii is the angle between the jet \pTivec and \bMHTivec, its
cosine is
\begin{align*}
  \cos\bDphii = \frac{\pTivec\cdot\left(\MHTvec +
    \pTivec\right)}{\left|\pTivec\right|\left|\MHTvec +
    \pTivec\right|},
\end{align*}
which can be written as
\begin{align}
  \label{eq:cosbdphi_f_dphi}
  \cos\bDphii = \frac{\fii +\cos\Dphii}{\sqrt{1 + \fii^2+2\fii\cos\Dphii}},
\end{align}
where \fii is the ratio of the magnitudes of \pTivec and \MHTvec:
\begin{align}
  \label{eq:fi}
  \fii \equiv \frac{\pTisca}{\MHTsca}.
\end{align}
This ratio plays a key role in this paper. In
Eq.~\eqref{eq:cosbdphi_f_dphi}, \bDphii is written as a function of
two dimensionless variables, \Dphii and \fii.
Figure~\ref{fig:f040_xxx_015_001} depicts the geometric relations
among \bDphii, \Dphii, and \fii in the transverse momentum
($p_x$-$p_y$) plane that is normalized as follows.

\paragraph{\boldmath Normalized \pTsca plane.}
The \textit{normalized $p^{}_T$ plane}---the coordinate plane in
Fig.~\ref{fig:f040_xxx_015_001}---is the transverse momentum plane
that is rotated and scaled such that the vector \MHTvec points
horizontally to the right with unit length. Moreover, the plane is
flipped if necessary so that \Dphii lies between 0 and $\pi$. The unit
vector \OCvec is \MHTvec in this plane. The vector \OAvec, with the
length \fii, is the jet \pTivec. The angle $\angle$AOC is, therefore,
the angle \Dphii. The normalized \pTsca plane is used throughout in
this paper.

\paragraph{Near side, away side.}
If the angle \Dphii is an acute angle, the jet is in the \textit{near
  side} of \MHTvec (Fig.~\ref{fig:f040_010_015_001})---the jet is a
near side jet; if the angle \Dphii is an obtuse angle, the jet is in
the \textit{away side} of \MHTvec
(Fig.~\ref{fig:f040_030_015_001})---the jet is an away side jet.

\bigbreak

The vector \OBvec is \bMHTivec. The length OB is \bMHTisca in the
scale of this plane:
\begin{align}
  \label{eq:length_ob}
  |\OBvec| &= \sqrt{1 + \fii^2+2\fii\cos\Dphii} \\ &=
  \frac{\sin\Dphii}{\sin\bDphii} \quad \text{(unless \Dphii is 0 or
    $\pi$)} \nonumber.
\end{align}
The angle $\angle$AOB is \bDphii, and so is its alternate angle
$\angle$OBC as shown in Fig.~\ref{fig:f040_xxx_015_001}.

\subsection{Analytic properties}
\label{subsec:analytic_properties}

\begin{figure}[!h]
\centering
\includegraphics[scale=1.0]{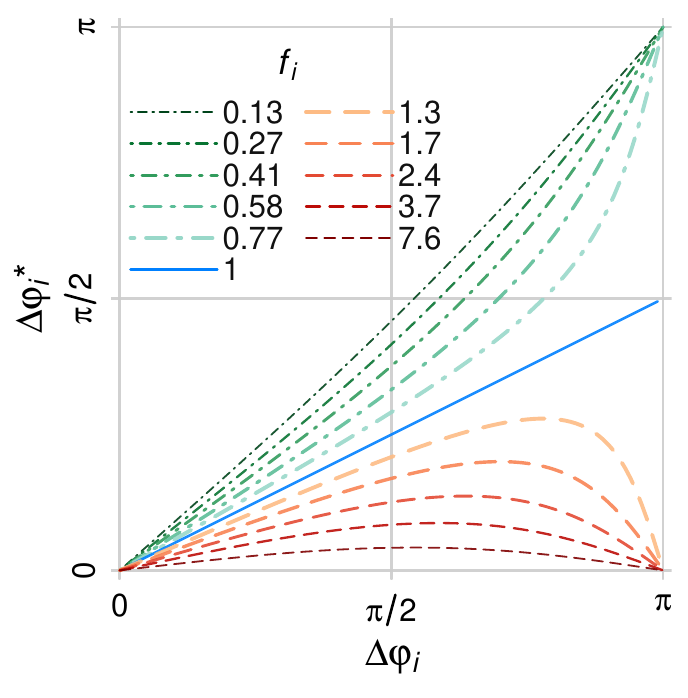}
\caption{\label{fig:f010_100_bdphi} The angle \bDphii as a function of
  the angle \Dphii for different values of \fii.}
\end{figure}

Equation~\eqref{eq:cosbdphi_f_dphi} relates \bDphii to $\Dphii$ and
$\fii$. The angle \bDphii is plotted in Fig.~\ref{fig:f010_100_bdphi}
as a function of \Dphii for different values of \fii. The angle
\bDphii is never wider than \Dphii. The angle \bDphii is zero when
\Dphii is zero for any values of \fii. In the limit of $\fii \to 0$,
\bDphii is the same as \Dphii in the entire range from 0 to $\pi$. For
any given value of \Dphii, the larger the ratio \fii, the narrower the
angle \bDphii. If $\fii < 1$, \bDphii is wider than half the angle
\Dphii and is the same as \Dphii at $\pi$. If $\fii = 1$, \bDphii is
half the angle \Dphii except when $\Dphii = \pi$, at which \bDphii is
indeterminate.\footnote{\label{foot:fone_dphipi} If $\fii = 1$ and
  $\Dphii = \pi$, \bDphii is indeterminate because then \bMHTivec is a
  zero vector, which is always the case for the jets in monojet
  events. If $\fii = 1$, as \Dphii approaches $\pi$, \bDphii
  approaches $\pi/2$. It is, therefore, sensible and convenient to
  define $\left.\bDphii\right|_{\Dphii=\pi, \fii=1} \equiv \pi/2$.} If
$\fii > 1$, \bDphii is narrower than half the angle \Dphii and is zero
when \Dphii is $\pi$. As \fii approaches infinity, \bDphii converges
to zero for any values of~\Dphii.

\section{\boldmath Minimizing \texorpdfstring{\MHTsca}{MHT} by varying jet \texorpdfstring{\pTsca}{pT}}
\label{sec:minimized_mht}

A jet mismeasurement or neutrinos in hadron decays in a jet can
largely alter the magnitude of the jet \pTvec from that of the
``true'' jet \pTvec (which we loosely define as the jet \pTvec at the
hadron level, before leptonic or semileptonic decays). However,
normally, the direction of the jet \pTvec is not largely altered.
Therefore, if we vary the magnitude of \pTvec of the jet that is
mismeasured or has neutrinos from hadron decays while retaining its
direction, the jet \pTvec becomes at some point in the variation close
to the ``true'' jet \pTvec.

Consequently, if large \MHTsca in a QCD multijet event is caused by a
jet mismeasurement or neutrinos in hadron decays in a jet, it will be
possible to make \MHTsca small by varying the \pTsca of one of the
jets in the event. In contrast, it is not generally possible to make
large genuine \MHTsca in a signal event small by varying the \pTsca of
any of the jets in the event.

In this section, we consider the variation of the magnitude of the jet
\pTivec and determine the minimized \MHTsca in the variation. The
range of the variation that we consider is from zero to infinity
rather than a range comparable to the jet \pTsca resolution because
large \MHTsca can be caused by a \textit{substantial} jet
mismeasurement or \textit{high-\pTsca} neutrinos in hadron decays. At
the end of the section, we show the relation of the minimized \MHTsca
to the angle \bDphii. In Section~\ref{sec:min_minimized_mht}, we will
demonstrate that the minimized \MHTsca for the jet that is mismeasured
or has neutrinos from hadron decays is indeed small in a QCD multijet
event with large \MHTsca in the simulated event samples.

\subsection{Minimized \texorpdfstring{\MHTvec}{MHT} and minimizing jet \texorpdfstring{\pTivec}{pT}}
\label{subsec:min_mht}

\begin{figure}[!b]
  \centering
  \begin{subfigure}[t]{0.48\textwidth}
    \centering
    \includegraphics[scale=0.85]{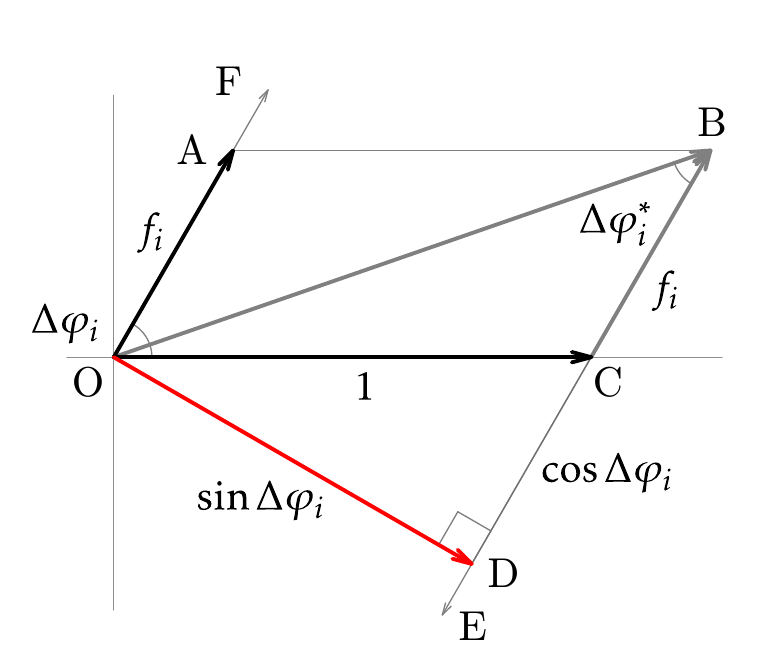}
    \caption{\label{fig:f040_010_020_001} $\fii + \cos\Dphii \ge 0$}
  \end{subfigure}
  ~
  \begin{subfigure}[t]{0.48\textwidth}
    \centering
    \includegraphics[scale=0.80]{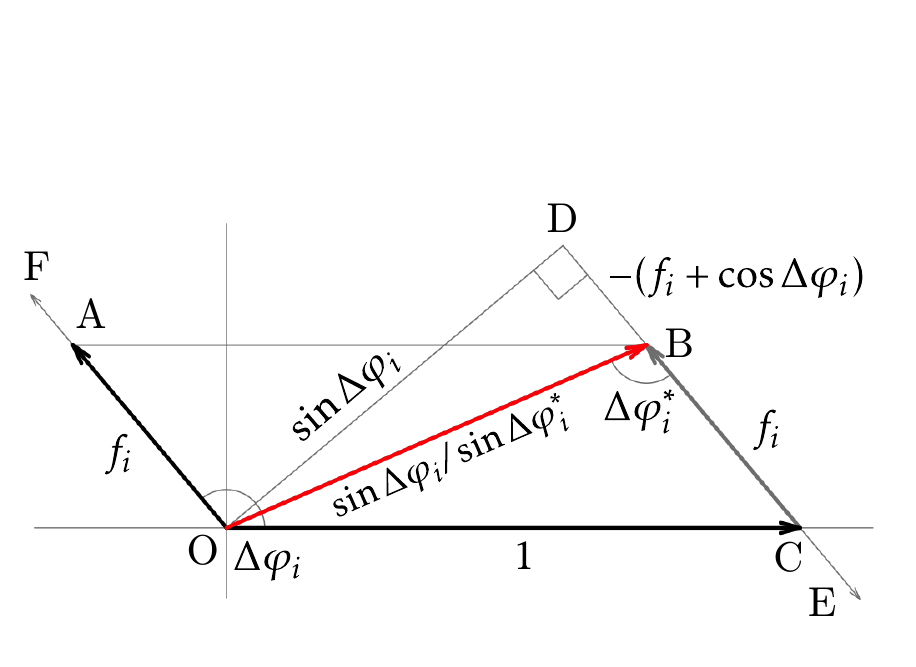}
    \caption{\label{fig:f040_070_020_001} $\fii + \cos\Dphii < 0$}
  \end{subfigure}
  \caption{\label{fig:f040_xxx_020_001} Minimized \MHTvec in the
    variation of the magnitude of the jet \pTivec in the normalized
    \pTsca plane, the coordinate plane used in
    Fig.~\ref{fig:f040_xxx_015_001}, in which the unit vector \OCvec
    is the initial \MHTvec and the vector \OAvec as well as \CBvec is
    the initial \pTivec. In the variation, the head of \MHTvec moves
    on the ray BE while the head of \pTivec on the ray OF. The vector
    \ODvec is the minimized \MHTvec in (a). In (b), \OBvec is the
    minimized \MHTvec as the point D is not on the ray~BE. }
\end{figure}

Let us consider the variation of the magnitude of the jet \pTivec and
determine the values of \MHTvec and \pTivec when \MHTvec is minimized
in the variation. We call them the \textit{minimized} \MHTvec and the
\textit{minimizing} jet \pTivec, respectively.

\paragraph{Notations.}
With regard to the notations in this discussion, the vectors \pTivec
and \MHTvec and their magnitudes \pTisca and \MHTsca change their
values in the variation, whereas \Dphii and \fii are constant with the
values determined by the initial \pTivec and initial \MHTvec. For
example, the angle between \pTivec and \MHTvec changes in the
variation, but \Dphii always denotes the angle between the initial
\pTivec and initial \MHTvec. Similarly, \fii always denotes the ratio
of the initial \pTisca and initial \MHTsca. Because \Dphii and \fii
are constant, the normalized \pTsca plane, the coordinate plane
introduced in Section~\ref{subsec:geo_bdphi_dphi_fii}, does not change
with the variation.

\bigbreak
From Eq.~\eqref{eq:bMHTivec}, \MHTvec is the vector difference of
\bMHTivec and \pTivec,
\begin{align}
  \label{eq:mht_bmht_pt}
  \MHTvec = \bMHTivec - \pTivec,
\end{align}
which initially corresponds to
\begin{align*}
  \OCvec = \OBvec - \CBvec
\end{align*}
in the normalized \pTsca plane, as can be seen in
Fig.~\ref{fig:f040_xxx_020_001}.

By definition, \bMHTivec does not change in the variation because its
value is determined by the other jets in the event---the vector \OBvec
is always \bMHTivec. Thus, Eq.~\eqref{eq:mht_bmht_pt} makes it clear
how \MHTvec changes when the magnitude of \pTivec is varied. If $\fii
+ \cos\Dphii \ge 0$, the vectors \ODvec and \DBvec are, respectively,
the minimized \MHTvec and the minimizing jet \pTivec
(Fig.~\ref{fig:f040_010_020_001}); if $\fii + \cos\Dphii < 0$, the
vector \OBvec is the minimized \MHTvec, and the minimizing jet \pTivec
is a zero vector~$\vec{0}$ (Fig.~\ref{fig:f040_070_020_001}).

\subsection{Scale factor \texorpdfstring{$\sin\DphiTildei$}{sin(Dphii\textasciitilde)} and ratio \texorpdfstring{\gii}{gi}}
\label{subsec:sin_dphi_tilde_g}

We have determined the two vectors in the normalized \pTsca plane: the
minimized \MHTvec and the minimizing jet \pTivec. Here, we will
introduce two dimensionless quantities---the scale factor
$\sin\DphiTildei$ and the ratio \gii---to denote the two vectors'
magnitudes in the scale of the normalized \pTsca plane.

\paragraph{\boldmath Scale factor $\sin\DphiTildei$.}
The scale factor $\sin\DphiTildei$ is the minimized \MHTsca in the
scale of the normalized \pTsca plane, which is the length~OD if $\fii
+ \cos\Dphii \ge 0$ and the length~OB (Eq.~\eqref{eq:length_ob})
otherwise\footnote{In this paper, the angle \DphiTildei always appears
  in the sine function. The angle \DphiTildei itself can be defined~as
\begin{align*}
\DphiTildei
&\equiv
\begin{cases}
    \Dphii & \text{if $\fii + \cos\Dphii \ge 0$}\\
    \pi - \arcsin\left(\sqrt{1 + \fii^2+2\fii\cos\Dphii}\right) & \text{otherwise}.
\end{cases}
\end{align*}
}:
\begin{align}
\label{eq:sin_dphi_tilde}
\sin\DphiTildei
&\equiv
\begin{cases}
    \sin\Dphii & \text{if $\fii + \cos\Dphii \ge 0$}\\
    \sqrt{1 + \fii^2+2\fii\cos\Dphii} & \text{otherwise}.
\end{cases}\\
&= \sqrt{1 +
   \left(\min\left(\fii, -\cos\Dphii\right)\right)^2+2\min\left(\fii, -\cos\Dphii\right)\cos\Dphii} \nonumber
\end{align}
In other words, $\sin\DphiTildei$ is the factor by which \MHTsca is
scaled from its initial value when it is minimized in the variation of
the jet \pTisca.

\bigbreak

\begin{figure}[!h]
\centering
\includegraphics[scale=1.0]{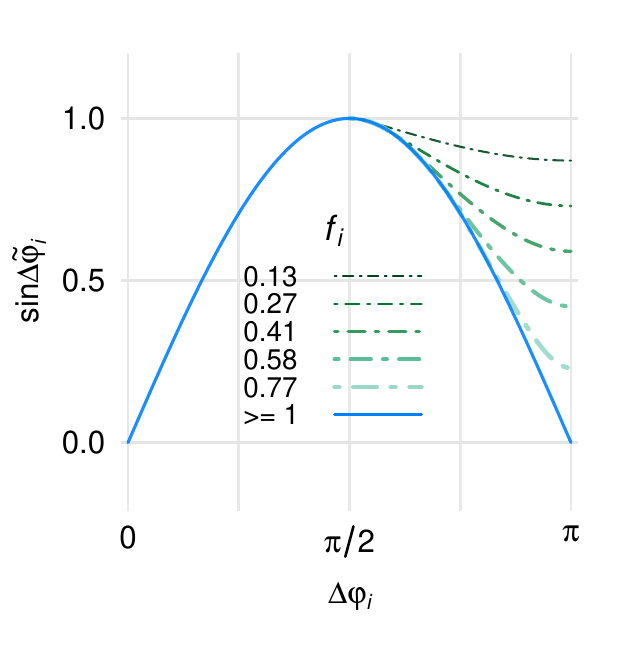}
\caption{\label{fig:f310_010} The scale factor $\sin\DphiTildei$ as a
  function of \Dphii for different values of \fii.}
\end{figure}

The scale factor $\sin\DphiTildei$ is plotted in
Fig.~\ref{fig:f310_010} as a function of \Dphii for different values
of~\fii. The scale factor $\sin\DphiTildei$ is larger than
$\sin\Dphii$ if $\fii + \cos\Dphii < 0$, that is, if \OBvec is the
minimized \MHTvec in the normalized \pTsca plane as in
Fig.~\ref{fig:f040_070_020_001}.

\paragraph{\boldmath Ratio \gii.}
The ratio $\gii$ is the minimizing jet \pTisca in the scale of the
normalized \pTsca plane, which is the length BD if $\fii + \cos\Dphii
\ge 0$ and is zero otherwise:
\begin{align}
\label{eq:gi}
\gii &\equiv
\begin{cases}
  \fii + \cos\Dphii & \text{if $\fii + \cos\Dphii \ge 0$}\\
  0 & \text{otherwise}
\end{cases}\\
&= \max(\fii + \cos\Dphii, 0). \nonumber
\end{align}
As a result, $\gii$ is the ratio of the minimizing jet \pTisca and the
initial \MHTsca.

\paragraph{\boldmath Minimized \MHTsca and minimizing jet \pTisca.}
In terms of $\sin\DphiTildei$ and \gii, the minimized \MHTsca and the
minimizing jet \pTisca are
\begin{align}
  \label{eq:minimized_mht}
  \begin{split}
  \text{minimized } \MHTsca &= \MHTsca\sin\DphiTildei\\
  \text{minimizing jet } \pTisca &= \gii\MHTsca,
  \end{split}
\end{align}
where \MHTsca in the right-hand side is the initial \MHTsca.

\bigbreak

\subsection{Relation to angle \texorpdfstring{\bDphii}{Dphii*}}
\label{subsec:alt_mean_bdphi}

We have determined the minimized \MHTsca as well as the minimizing jet
\pTisca. Before closing this section, we point out their relation to
the angle \bDphii. The ratio of minimized \MHTsca and the minimizing
jet \pTisca is
\begin{align}
  \label{eq:minmht_minpt}
  \frac{\text{minimized } \MHTsca}{\text{minimizing jet } \pTisca}
  = \frac{\sin\DphiTildei}{\gii} =
  \begin{cases}
    \displaystyle
    \frac{\sin\Dphii}{\fii + \cos\Dphii} & \text{if $\fii + \cos\Dphii \ge 0$}\\
    \infty & \text{otherwise}.
\end{cases}
\end{align}
From Eq.~\eqref{eq:cosbdphi_f_dphi}, $\tan\bDphii$ can be written as
\begin{align}
  \label{eq:tan_bdphii}
  \tan\bDphii = \frac{\sin\Dphii}{\fii + \cos\Dphii},
\end{align}
which is the same as the ratio for the case of $\fii + \cos\Dphii \ge
0$ in Eq.~\eqref{eq:minmht_minpt}. Therefore, in words, the relation
is that $\tan\bDphii$ is the ratio of the minimized \MHTsca and
minimizing jet \pTisca provided $\fii + \cos\Dphii \ge 0$. In
Sections~\ref{sec:altvars}, we will define alternative angles based on
similar relations.

\bigbreak

We started this section by arguing that it is possible to make large
\MHTsca in a QCD multijet event small by varying the \pTsca of one of
the jets in the event. We then determined the minimized \MHTsca in the
variation of a jet \pTsca. The minimized \MHTsca can be defined for
each jet in the event. In Section~\ref{sec:min_minimized_mht}, we will
show that the minimum of the minimized \MHTsca for all jets in a QCD
multijet event with large \MHTsca is indeed generally small in
simulated event samples, which will be described in the next section.

\section{\boldmath Simulated event samples}
\label{sec:sample}

We have generated Monte Carlo simulated events of proton-proton
collisions at $\sqrt{s}=13$~TeV. This section describes the simulated
event samples.

\paragraph{\boldmath MadGraph5+Pythia8.}
Events were generated at the parton level at leading order with
\MGvATNLO 2.3.3~\cite{Alwall:2014hca} with the parton distribution
function NNPDF2.3 LO~\cite{Ball:2012cx}. The fragmentation and parton
shower were simulated by \PYTHIA 8.2 \cite{Sjostrand:2014zea} with the
MLM matching~\cite{Alwall:2007fs}.


\paragraph{\boldmath Background processes.}
We generated events for the following standard model processes as
background samples.

\begin{description}[leftmargin=\parindent,labelindent=\parindent]
\item[QCD.] QCD multijet events were generated with up to three
  outgoing partons in the matrix element calculation of
  \MGvATNLO.\footnote{We compared with a small sample generated with
    up to four outgoing partons and observed agreement in
    distributions of the leading jet \pTsca, the inclusive jet \pTsca,
    the jet multiplicity, \HT, \MHTsca, and \METsca.}
  \item[EWK.] In this paper, \ttjets, \wjets, and \znnjets events are
    collectively called EWK events, as done for example in
    Ref.~\cite{Khachatryan:2011tk}. Apart from QCD multijet events,
    these processes are typical dominant background processes in
    all-hadronic SUSY searches as for example in
    Refs.~\cite{Sirunyan:2017cwe, Aaboud:2017hrg}. The \ttjets events
    were generated with up to three additional outgoing partons.
    \MADSPIN~\cite{Artoisenet:2012st} was used for the top quark decay
    in the \ttjets events. The \wjets and \znnjets events were
    generated with up to four additional outgoing partons.
\end{description}

\paragraph{\boldmath Signal processes.}

As benchmark signal models, we generated events in two models from a
class of simplified SUSY models~\cite{Alwall:2008ag, Alves:2011wf} in
which gluinos are produced in pairs and each gluino decays into a
top-antitop quark pair and a neutralino as the LSP
($\PSg\to\ttbar\PSGczDo$). This class of models is called T1tttt in
CMS~\cite{Chatrchyan:2013sza} and Gtt in ATLAS~\cite{Aad:2012pq}.
Models are fully specified when the gluino mass $m_\PSg$ and the
neutralino mass $m_\PSGczDo$ are specified. We picked two pairs of the
masses and refer the two signal models as follows:
\begin{description}[leftmargin=\parindent,labelindent=\parindent]
  \item[T1tttt(1950,~500).] $m_\PSg = 1950$~GeV and $m_\PSGczDo =
    500$~GeV, a \textit{high-mass gluino model} with large split
    between $m_\PSg$ and $m_\PSGczDo$.
  \item[T1tttt(1350,~1100).] $m_\PSg = 1350$~GeV and $m_\PSGczDo =
    1000$~GeV, a \textit{compressed spectrum}, in which the difference
    between $m_\PSg$ and $m_\PSGczDo$ is small.
\end{description}
These mass pairs are near the exclusion contour in the
$m_\PSg$-$m_\PSGczDo$ plane in this class of the models in recent CMS
and ATLAS results~\cite{Sirunyan:2017uyt, Sirunyan:2017cwe,
  Sirunyan:2017kqq, Sirunyan:2017fsj, Aaboud:2017hdf, Aaboud:2017dmy,
  Sirunyan:2017mrs, Sirunyan:2017hvp, Sirunyan:2017pjw,
  Aaboud:2017hrg, Sirunyan:2018vjp}. The events were generated with up
to two additional outgoing partons.

\paragraph{\boldmath Delphes with CMS card.}
The detector responses to the events were simulated by \DELPHES 3.4.1
\cite{deFavereau:2013fsa} with the CMS detector configuration card. We
used the configuration file \texttt{delphes\_card\_CMS\_PileUp.tcl}
included in the \DELPHES package with a slight
modification.\footnote{We changed the average number of pileup
  interactions and the distance parameter of the anti-$k_{\mathrm{T}}$
  algorithm and excluded neutralinos from generated jets.}

On average, 23 pileup interactions were placed in each event by
\DELPHES. Further, \DELPHES performed (a) the reconstruction of
physics objects such as jets, electrons, muons, photons, and missing
transverse momentum; (b) the calculations of the isolation variables
for electrons, muons, and photons; (c) the $\tau$ tagging of jets; and
(d) pileup subtraction on jets.

\paragraph{\boldmath Jets, generated jets, jet \pTsca correction.}
Jets were defined by the anti-$k_{\mathrm{T}}$
algorithm~\cite{Cacciari:2008gp} with the distance parameter 0.4 by
\FASTJET~\cite{Cacciari:2011ma, Cacciari:2005hq} within \DELPHES. In
addition, \textit{generated} jets as sets of particles after the
fragmentation, parton shower, and decay of certain short-lived
particles were defined also by the anti-$k_{\mathrm{T}}$ algorithm
with the distance parameter 0.4 by \FASTJET within \DELPHES. Generated
jets do not include neutrinos or neutralinos.

The jet transverse momenta \pTsca were corrected such that the peak
locations of the distributions of jet \pTsca agreed with those for the
generated jet \pTsca in ranges of \pTsca and the pseudorapidity
($\eta$). Jets with $\pTsca \ge 30$~GeV were used. The jet \pTsca
resolution is shown in Appendix~\ref{appendix:jer}.

\begin{figure}[!b]
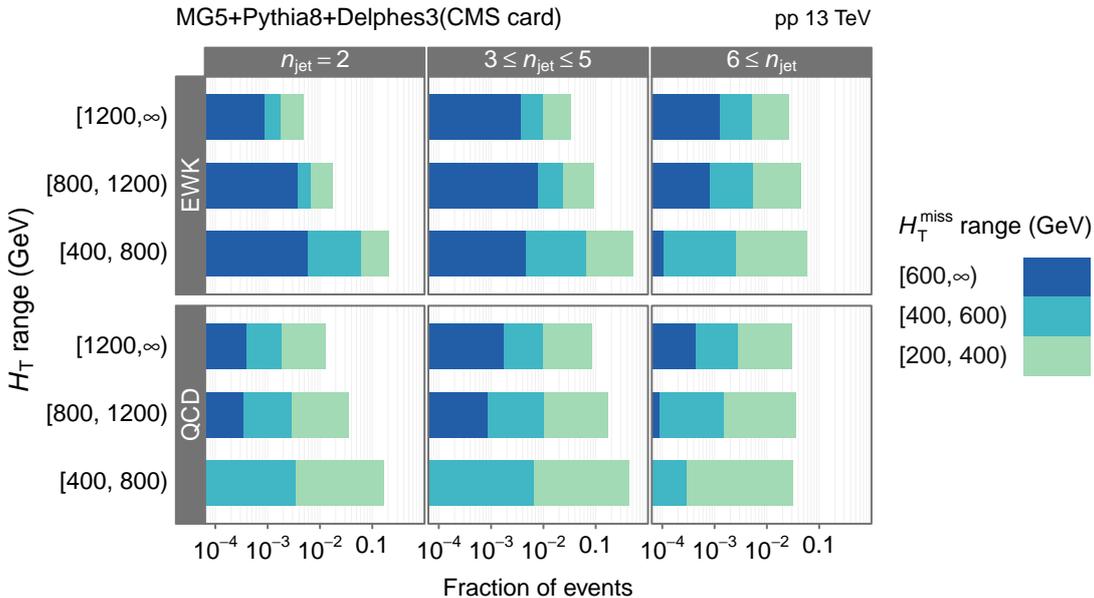

\centering
\includegraphics[scale=1.0]{{{g0220.barchart_ratio_mhtbin_njetbin_htbin.delphes_cms}}}
\caption{\label{fig:g0220.delphes_cms} The fractions of the events in
  ranges of \njet, \HT, and \MHTsca for the background processes after
  the event selection is applied. The fractions add up to unity
  separately for the QCD and EWK events.}
\end{figure}

\begin{figure}[!t]
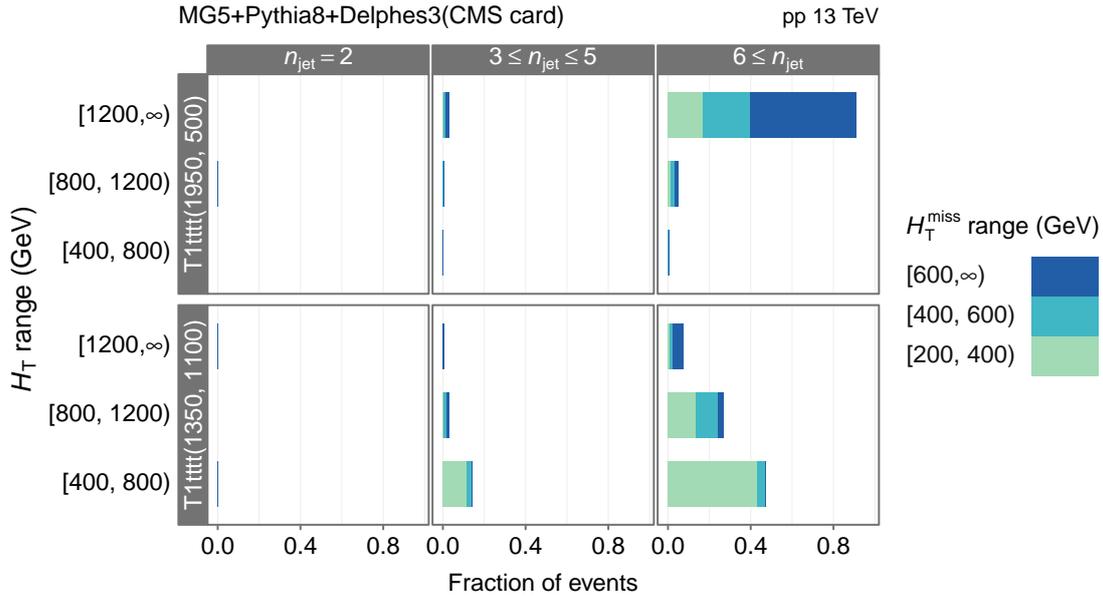

\centering
\includegraphics[scale=1.0]{{{g0210.barchart_ratio_mhtbin.njetbin_htbin_signal.delphes_cms}}}
\caption{\label{fig:g0210.delphes_cms} The fractions of the events in
  ranges of \njet, \HT, and \MHTsca for the two signal models,
  T1tttt(1950,~500) and T1tttt(1350,~1100), after the event selection
  is applied. The fractions add up to unity separately for each
  model.}
\end{figure}

\paragraph{\boldmath Event selection.}
Events with any of at least one isolated electron, isolated muon, or
isolated photon, were discarded. Events were discarded if they contain
a jet that did not satisfy quality criteria\footnote{In terms of
  variables in \DELPHES, the criteria are $\texttt{Beta} \ge 0.14$,
  $\texttt{NCharged} > 0$, $\texttt{PTD} < 0.8$, and
  $\texttt{MeanSqDeltaR} < 0.1$.} or that was tagged as hadronically
decaying $\tau$.

Events were required (a) to have at least two jets ($\njet \ge 2$);
(b) to satisfy $\HT \ge 400$~GeV where \HT is the scalar sum of \pTsca
of all jets in the event; (c) to have large \MHTsca, namely, $\MHTsca
\ge 200$~GeV; and (d) to satisfy $\MHTsca/\METsca < 1.25$, which, as
used for example in Ref.~\cite{Khachatryan:2016dvc}, ensures that the
large \MHTsca is not due to unclustered objects or jets with \pTsca
below the threshold of 30~GeV.

\bigbreak

The fractions of the selected events in ranges of \njet, \HT, and
\MHTsca for the QCD and EWK events are shown in
Fig.~\ref{fig:g0220.delphes_cms} and for the two signal models in
Fig.~\ref{fig:g0210.delphes_cms}. High jet multiplicity events ($\njet
\ge 6$) dominate for both signal models. Most events in
T1tttt(1950,~500) are in the highest-\HT range. A large fraction of
events in T1tttt(1350,~1100) have low to medium \HT and low \MHTsca.

\section{\boldmath Minimized \texorpdfstring{\MHTsca}{MHT} in QCD multijet events}
\label{sec:min_minimized_mht}

At the beginning of Section \ref{sec:minimized_mht}, we argued that
the minimized \MHTsca in the variation of \pTsca of one of the jets in
a QCD multijet event with large \MHTsca should be small as long as the
large \MHTsca is caused by a jet mismeasurement or neutrinos in hadron
decays in a jet, while in a signal event the minimized \MHTsca is not
necessarily small for any of the jets in the event. In the simulated
event samples just described in the previous section, this section
confirms the argument.

\begin{figure}[!b]
\centering
\includegraphics[scale=1.0]{{{g0320.xyplot.log10.val.mht_htbin.process.htbin_800.var_010.delphes_cms}}}
\caption{\label{fig:g0320.delphes_cms} The distributions of QCD, EWK,
  T1tttt(1350,~1100), and T1tttt(1950,~500) events in \MHTsca and
  three related variables: \mmMHTsca, defined in
  Eq.~\eqref{eq:minimum_minimized_mht}; the gen \MHTsca, the \MHTsca
  based on generated jets, which do not include neutrinos; the
  ``true'' \MHTsca, the \MHTsca based on the generated jets including
  the neutrinos within $\Delta R = \sqrt{\Delta\varphi^2 +
    \Delta\eta^2} \le 0.4$ from the axis of any of the generated jets.
  Each distribution is normalized to unity. }
\end{figure}

We let \mmMHTsca denote the minimum among the minimized \MHTsca for
all jets in the event:
\begin{align}
  \label{eq:minimum_minimized_mht}
  \mmMHTsca \equiv \MHTsca \min_{i\in\text{jets}}\sin\DphiTildei,
\end{align}
where $\sin\DphiTildei$ is defined in Eq.~\eqref{eq:sin_dphi_tilde}.

Figure~\ref{fig:g0320.delphes_cms} shows that, in the simulated event
samples, the distribution of QCD multijet events in \mmMHTsca quickly
decreases by orders of magnitude while the distributions in \mmMHTsca
for the other processes decrease more slowly: \mmMHTsca is indeed
generally small in QCD events with large \MHTsca. Moreover, the figure
also shows that, in QCD multijet events, \mmMHTsca roughly agrees with
``true'' \MHTsca, which is estimated by the magnitude of the vector
sum of \pTvec of the generated jets and the neutrinos within $\Delta R
= \sqrt{\Delta\varphi^2 + \Delta\eta^2} \le 0.4$ from the axis of any
of the generated jets.

The small \mmMHTsca in QCD multijet events with large \MHTsca and its
rough agreement with the ``true'' \MHTsca indicate (a) that the large
\MHTsca in QCD multijet events is in fact dominantly caused by a jet
mismeasurement and neutrinos in hadron decays in a jet and (b) that,
in QCD multijet events with large \MHTsca, the minimizing \pTsca for
the jet that yields \mmMHTsca is in fact close to the ``true''
jet~\pTsca.

\bigbreak

This short section has confirmed the argument made at the beginning of
Section \ref{sec:minimized_mht} in simulated event samples. This
confirmation suggests that \mmMHTsca can possibly replace \MHTsca (or
\METsca) in the event selection, which is pursued in
Appendix~\ref{appendix:mmmht_event_selection}.

\section{\boldmath Features of \texorpdfstring{\minbDphi}{minDphi*} criterion}
\label{sec:min_bdphi_feature}

This section reviews the features of the \minbDphi criterion. Let us
first restate the definitions of the angular variable \minbDphi and
the \minbDphi criterion.

\paragraph{\boldmath Angular variable \minbDphi.}
The angular variable \minbDphi is the minimum of the angles \bDphii,
defined in \eqref{eq:bdphi}, for all jets in the event:
\begin{align}
  \label{eq:min_bdphi}
  \minbDphi \equiv \min_{i\in\text{jets}}\bDphii.
\end{align}

\paragraph{\boldmath The \minbDphi criterion.}
The \minbDphi criterion of the event selection requires events to have
\minbDphi above a threshold $\gamma_0$:
\begin{align}
  \label{eq:minbdphi_ge_gamma}
  \minbDphi \ge \gamma_0.
\end{align}
In other words, the \minbDphi criterion rejects any event with at
least one jet with the angle \bDphii narrower than $\gamma_0$. The
value of $\gamma_0$ is 0.5 in Refs.~\cite{Khachatryan:2016dvc,
  Sirunyan:2018vjp}. The discussion in this section is not valid if
$\gamma_0$ is too large, for example, as large as $\pi/2 \approx 1.57$
or larger.

\bigbreak

Section~\ref{subsec:analytic_properties} reviewed the properties of
\bDphii as a function of \Dphii and \fii. Here, we will relate the
properties with features of the \minbDphi criterion. In particular, we
consider large \fii, wide \Dphii, and narrow \Dphii and identify room
for improvement.

\subsection{Large \texorpdfstring{\fii}{fi} and \texorpdfstring{\maxf}{maxf}}
\label{subsec:large_f}

When \fii is larger than unity, \bDphii is not an increasing function
of \Dphii as can be seen in Fig.~\ref{fig:f010_100_bdphi}. From
Eq.~\eqref{eq:cosbdphi_f_dphi}, \bDphii takes its maximum value of
$\arcsin(1/\fii)$ when $\cos\Dphii = -1/\fii$. For example, since
$1/\sin\gamma_0|_{\gamma_0 = 0.5} = 2.09$, the \minbDphi criterion
with $\gamma_0 = 0.5$ rejects any event with a jet with \pTsca at
least 2.09 times as large as \MHTsca.

\begin{figure}[!h]
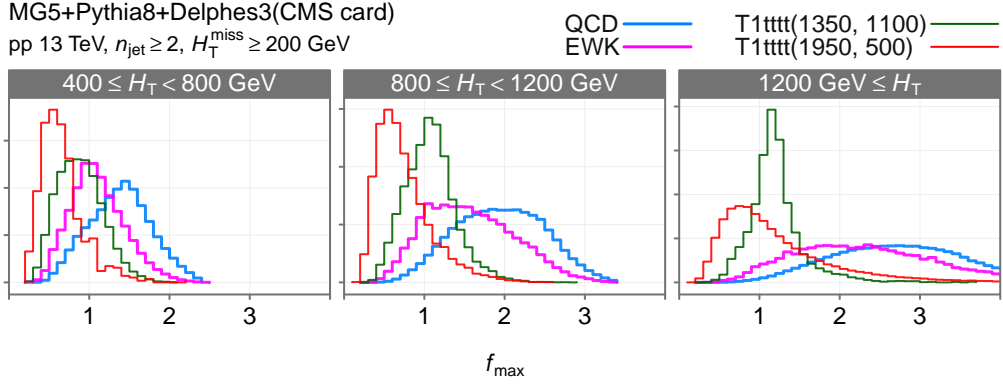

\centering
\includegraphics[scale=1]{{{g0110_520.n.process.maxF.delphes_cms}}}
\caption{\label{fig:g0110_520.n.process.maxF.delphes_cms} The
  distributions of QCD, EWK, T1tttt(1950,~500), and T1tttt(1350,~1100)
  events in \maxf in three ranges of \HT in the simulated samples
  described in Section~\ref{sec:sample}. Each distribution is
  normalized to unity.}
\end{figure}

In general, the \minbDphi criterion rejects all events with
\begin{align}
  \label{eq:maxf_sin_gamma}
  \maxf \equiv \max_{i\in\text{jets}}\fii \ge
  \frac{1}{\sin\gamma_0}.
\end{align}
This feature might appear to needlessly reduce signal acceptances; on
the contrary, it effectively reduces background events without much
decreasing signal acceptances. In fact, the distributions of QCD and
EWK background events in \maxf have the peaks at larger values than
those of signal events, as shown in
Fig.~\ref{fig:g0110_520.n.process.maxF.delphes_cms}.\footnote{The
  variable \maxf could, therefore, be useful for offline and online
  preselection of events for reducing data sizes and event rates.}

\clearpage 

\subsection{Wide \texorpdfstring{\Dphii}{Dphii} and jet \texorpdfstring{\pTsca}{pT} overestimate}
\label{subsec:large_dphi}

\begin{figure}[!b]
  \begin{minipage}[c]{0.45\textwidth}
    \includegraphics[scale=1]{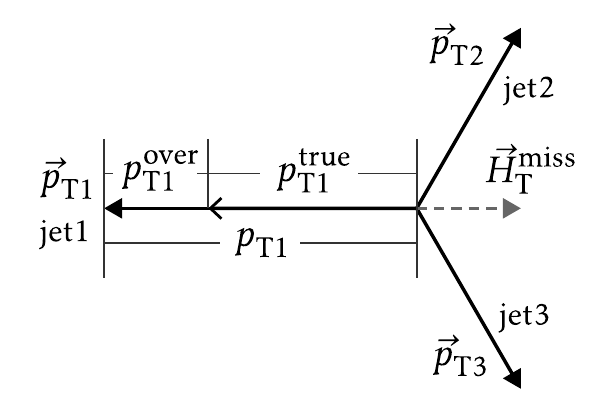}
  \end{minipage}\hfill
  \begin{minipage}[c]{0.53\textwidth}
    \caption{\label{fig:qcd_ overmeasurement} Schematic illustrating
      when a jet \pTsca overestimate causes large \MHTsca in a QCD
      event, the jet \pTvec and \MHTvec are back-to-back and the jet
      \pTsca is greater than \MHTsca. The jet \pToneSca, with its true
      value \pToneTrueSca, is overestimated by \pToneOverSca, which
      causes \MHTvec. The \MHTsca corresponds to the amount by which
      \pToneSca is overestimated, i.e., $\MHTsca = \pToneOverSca$;
      therefore, $\pToneSca > \MHTsca$.}
  \end{minipage}
\end{figure}
\begin{figure}[!t]
  \begin{minipage}[c]{0.45\textwidth}
    \includegraphics[scale=1]{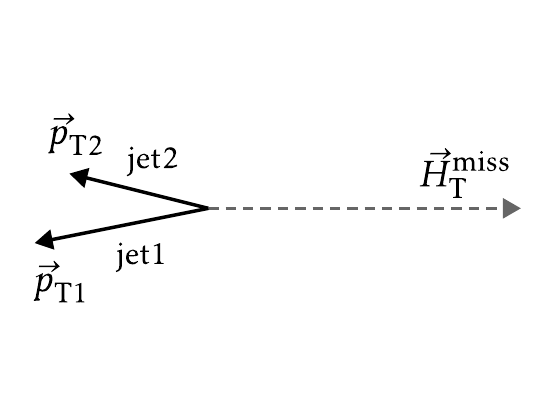}
  \end{minipage}\hfill
  \begin{minipage}[c]{0.53\textwidth}
    \caption{\label{fig:susy_recoil} A simple example of signal events
      where jets back-to-back with \MHTvec have each \pTsca smaller
      than \MHTsca. Because the two jets share the recoil of \MHTvec,
      each jet has \pTisca ($i = 1, 2$) smaller than \MHTsca.}
  \end{minipage}
\end{figure}

Consider wide \Dphii (near $\pi$). The jet and \MHTvec are
back-to-back. As can be seen in Fig.~\ref{fig:f010_100_bdphi}, when
\Dphii is near $\pi$, \bDphii will be below the threshold $\gamma_0$
only if $\fii > 1$. Consequently, the \minbDphi criterion rejects
every event with a jet back-to-back with \MHTvec if the jet \pTsca is
greater than \MHTsca, whereas the criterion does not reject any event
with a jet back-to-back with \MHTvec if the jet \pTsca is less than
\MHTsca unless another jet in the same event has narrow \bDphii. This
difference based on whether the jet \pTsca is greater or less than
\MHTsca is an advantage in rejecting QCD multijet background events
with large \MHTsca caused by a jet \pTsca \textit{overestimate}.

If a jet \pTsca overestimate causes large \MHTsca, the jet and \MHTvec
are back-to-back and the jet \pTsca is greater than \MHTsca. The
reason why the jet \pTsca is greater than \MHTsca is that \MHTsca
corresponds only to the amount by which the jet \pTsca is
overestimated. An example is illustrated in Fig.~\ref{fig:qcd_
  overmeasurement}. In contrast, in signal events where large \MHTsca
is due to invisible particles, a jet back-to-back with \MHTvec does
not necessarily have \pTsca greater than \MHTsca. A simple example is
given in Fig.~\ref{fig:susy_recoil}.

\subsection{Narrow \texorpdfstring{\Dphii{}}{Dphii}---room for improvement}
\label{subsec:small_dphi}

Consider narrow \Dphii. If large \MHTsca in a QCD multijet event is
caused by a jet \pTsca \textit{underestimate}, which can be due to
either a mismeasurement or neutrinos in hadron decays, the jet has
narrow \Dphii. Because \bDphii is always narrow when \Dphii is narrow,
the \minbDphi criterion rejects QCD multijet background events with
large \MHTsca caused by a jet \pTsca underestimate.

As a function of \Dphii and \fii, \bDphii is larger for smaller \fii
for a given value of \Dphii (Fig.~\ref{fig:f010_100_bdphi}).
Therefore, the \minbDphi criterion, in effect, lowers the threshold on
\Dphii of a jet with smaller \fii. This is advantageous because, in a
QCD multijet event with large \MHTsca caused by a jet \pTsca
underestimate, \Dphii of the jet whose \pTsca is underestimated tends
to be narrower if its \fii is smaller. The reason is as follows: the
smaller \fii indicates that a larger fraction of the ``true'' jet
\pTsca contributes to \MHTsca, aligning \MHTvec more with the ``true''
jet \pTvec, which in turn results in narrower~\Dphii, as long as the
jet \pTivec is parallel to the ``true'' jet~\pTivec.\footnote{In
  general, the jet \pTivec is not exactly parallel to the ``true'' jet
  \pTivec. In fact, the larger the underestimate, the more deflected
  the jet \pTvec is likely to be from the direction of the ``true''
  jet \pTvec. This is a minor counter effect to the described relation
  between \Dphii and \fii.}

However, because \bDphii is never wider than \Dphii by definition, the
effective threshold of \Dphii is never lower than $\gamma_0$.
Consequently, the \minbDphi criterion rejects every event with at
least one jet with $\Dphii < \gamma_0$. This feature needlessly
reduces signal acceptances, especially of models with high jet
multiplicity, such as T1tttt, the benchmark signal models used in this
paper. There is room for improvement here. For example, it is possible
to keep signal acceptances larger if the effective threshold of \Dphii
can be arbitrarily low for sufficiently small \fii. Angular variables
with such a property will be introduced in the next section.

\section{\boldmath Alternative angular variables: \texorpdfstring{\minOmegaTilde}{minomega{\textasciitilde}}, \texorpdfstring{\minOmegaHat}{minomega{\textasciicircum}}, \texorpdfstring{\minChi}{minchi}}
\label{sec:altvars}

In the previous section, we have reviewed the features of the
\minbDphi criterion and identified room for improvement. In this
section, we introduce three alternative angular
variables---\minOmegaTilde, \minOmegaHat, and \minChi---after
introducing the angle \omegai and its three variants \omegaTildei,
\omegaHati, and~\chii.

\subsection{Angle \texorpdfstring{\omegai}{omegai}}
\label{subsec:omegai}

The angle \omegai brings the improvement for near side jets mentioned
at the end of the previous section. However, \omegai loses the
advantage of \bDphii for away side jets described in
Section~\ref{subsec:large_dphi}. Later, we will introduce three
variants of \omegai that recover the lost advantage.

We saw in Section~\ref{subsec:alt_mean_bdphi} that $\tan\bDphii$ is
the ratio of the minimized \MHTsca and the minimizing jet \pTisca as
long as $\fii + \cos\Dphii \ge 0$. The tangent of the angle \omegai is
a similar ratio.

\paragraph{\boldmath Angle \omegai.}
The angle \omegai is defined such that $\tan\omegai$ is the ratio of
$\sin\Dphii$ and \fii:
\begin{align}
  \label{eq:tan_omega}
  \tan\omegai \equiv \frac{\sin\Dphii}{\fii}.
\end{align}
The angle \omegai ranges in $0~\le \omegai < \pi/2$. The $\tan\omegai$
is the ratio of the minimized \MHTsca and the initial jet \pTisca if
$\fii + \cos\Dphii \ge 0$.

\begin{figure}[!b]
  \centering
  \begin{subfigure}[t]{0.48\textwidth}
    \centering
    \includegraphics[scale=0.85]{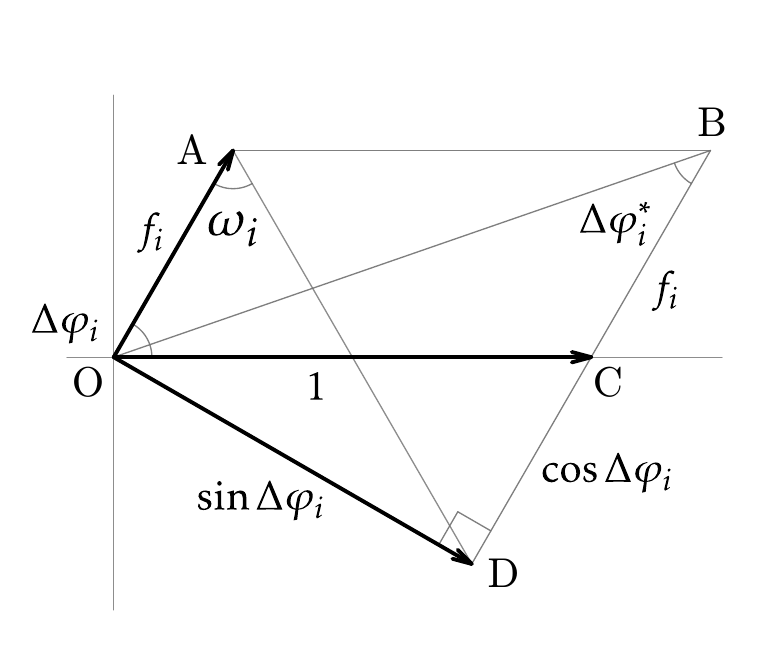}
    \caption{\label{fig:f040_010_030_001} $\Dphii < \pi/2$}
  \end{subfigure}
  ~
  \begin{subfigure}[t]{0.48\textwidth}
    \centering
    \includegraphics[scale=0.80]{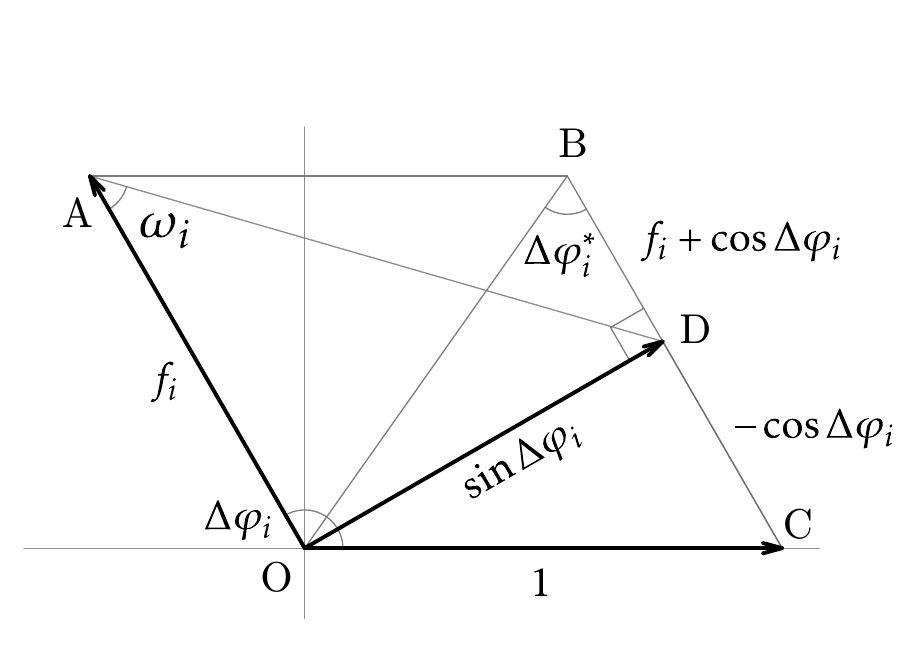}
    \caption{\label{fig:f040_030_030_001} $\Dphii > \pi/2$}
  \end{subfigure}
\caption{\label{fig:f040_xxx_030_001} The angle \omegai in the
  normalized \pTsca plane, where \OCvec and \OAvec are \MHTvec and the
  jet \pTivec, respectively. The angle $\angle$OAD is the angle
  \omegai.}
\end{figure}

\paragraph{Geometric relations.}
The angle \omegai appears in the normalized \pTsca plane as shown in
Fig.~\ref{fig:f040_xxx_030_001}: the angle $\angle$OAD is \omegai. The
angle \omegai is wider than \bDphii for a near side jet
(Fig.~\ref{fig:f040_010_030_001}) and is narrower than \bDphii for an
away side jet (Figs.~\ref{fig:f040_030_030_001}). If $\Dphii = \pi/2$,
the two angles are the same.

\begin{figure}[!h]
  \centering
  \includegraphics[scale=1.0]{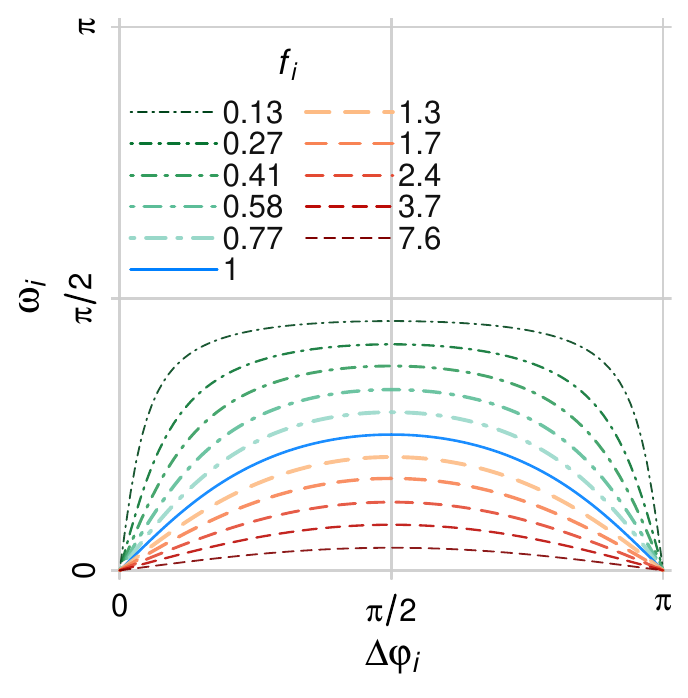}
  \caption{\label{fig:f010_200_omega} The angle \omegai as a
    function of the angle \Dphii for different values of \fii.}
\end{figure}

\paragraph{\boldmath As function of \Dphii and \fii.}
The angle \omegai is plotted as a function of \Dphii for different
values of \fii in Fig.~\ref{fig:f010_200_omega}. The function is
symmetric around $\Dphii = \pi/2$, at which it takes its maximum value
of $\arccot\fii$. The angle \omegai is zero when \Dphii is zero or
$\pi$. The smaller the ratio \fii, the more quickly the angle \omegai
increases from zero near $\Dphii = 0$ and decreases to zero near
$\Dphii = \pi$. In the limit of $\fii \to 0$, \omegai becomes a step
function of \Dphii that jumps from zero to $\pi/2$ at $\Dphii = 0$ and
from $\pi/2$ to zero at $\Dphii = \pi$.

\paragraph{\boldmath Improvement for near side jets.}
The improvement for near side jets discussed in
Section~\ref{subsec:small_dphi} can be seen in
Fig.~\ref{fig:f010_200_omega}. The angle \omegai can be wider than
\Dphii for near side jets. In fact, no matter how narrow the angle
\Dphii is (except 0), \omegai can be wider than any given acute angle
if \fii is sufficiently small. Furthermore, no matter how small the
ratio \fii is, \omegai can be narrower than any given angle if \Dphii
is sufficiently small. These properties let the near side jets whose
\pTsca are underestimated in QCD multijet events with large \MHTsca
have narrow \omegai while preventing any near side jet in signal
events from unnecessarily having narrow \omegai.

\paragraph{\boldmath Lost advantage for away side jets.}
The angle \omegai does not have the advantage of \bDphii for away side
jets described in Section~\ref{subsec:large_dphi}: the angle \omegai
is zero at $\Dphii = \pi$ regardless of whether \fii is greater or
less than unity. This property makes the angles \omegai of jets in
signal events unnecessarily narrow if the jets have \pTsca less than
\MHTsca and are back-to-back with~\MHTvec.

\subsection{Variants of \texorpdfstring{\omegai}{omegai}}
\label{subsec:variants_omegai}

We introduce three variants of the angle \omegai to recover the lost
advantage. These variants, denoted by \omegaTildei, \omegaHati, and
\chii, have the same value as \omegai for near side jets but can be
wider than \omegai for away side jets.

\subsubsection{Angle \texorpdfstring{\omegaTildei}{omega{\textasciitilde}i}}

The $\tan\omegai$ is the ratio of the minimized \MHTsca and the
initial jet \pTisca only if $\fii + \cos\Dphii \ge 0$. The first
variant, denoted by \omegaTildei, is defined such that its tangent is
always the ratio regardless of the sign of the sum $\fii +
\cos\Dphii$.

\paragraph{\boldmath Angle \omegaTildei.}
The angle \omegaTildei is defined such that $\tan\omegaTildei$ is the
ratio of $\sin\DphiTildei$ and \fii:
\begin{align}
  \label{eq:tan_omega_tilde}
  \tan\omegaTildei \equiv \frac{\sin\DphiTildei}{\fii},
\end{align}
where $\sin\DphiTildei$ is the scale factor defined in
Eq.~\eqref{eq:sin_dphi_tilde}. The angle \omegaTildei ranges in $0~\le
\omegaTildei < \pi/2$.

\bigbreak

Compared to $\tan\omegai$ in Eq.~\eqref{eq:tan_omega}, $\sin\Dphii$ in
the numerator is replaced with $\sin\DphiTildei$. The angle
\omegaTildei is plotted as a function of \Dphii for different values
of \fii in the leftmost panel of Fig.~\ref{fig:f015}. The angle
\omegaTildei is wider than \omegai if $\fii + \cos\Dphii < 0$ and is
not zero at $\Dphii = \pi$ if $\fii < 1$.

\subsubsection{Angle \texorpdfstring{\omegaHati}{omega{\textasciicircum}i}}
\label{subsubsec:omega_hati}

Consider the minimized \MHTsca in the variation of \pTisca of only
near side jets. The minimized \MHTsca in such variation can be
expressed as the following scale factor.

\paragraph{\boldmath Scale factor $\sin\DphiHati$.}
The scale factor $\sin\DphiHati$ is defined as
\begin{align}
\label{eq:sin_dphi_hat}
\sin\DphiHati
&\equiv
\begin{cases}
    \sin\Dphii & \text{if $\Dphii \le \pi/2$}\\
    1 & \text{otherwise}.
\end{cases}
\end{align}
The angle \DphiHati can be written as $\DphiHati = \min(\Dphii,
\pi/2)$.

\bigbreak

This scale factor is motivated by considering event topologies where
all jets in the event are away side jets, which are unlike topologies
for QCD multijet events and possible topologies for signal events. QCD
multijet events usually have both near and away side jets. With this
scale factor, \MHTsca is not minimized if the event has only away side
jets. When large \MHTsca in QCD multijet events is caused by a jet
\pTisca overestimate, the variation of the jet \pTisca that is
actually overestimated is not considered. But the variation of \pTisca
of near side jets in the same event can still minimize \MHTsca to some
extent.

\paragraph{\boldmath Angle \omegaHati.}
The angle \omegaHati is defined such that $\tan\omegaHati$ is the
ratio of $\sin\DphiHati$ and \fii:
\begin{align}
  \label{eq:tan_omega_hat}
  \tan\omegaHati \equiv \frac{\sin\DphiHati}{\fii}.
\end{align}
The angle \omegaHati ranges in $0 \le \omegaHati < \pi/2$. The
$\tan\omegaHati$ is the ratio of the minimized (initial) \MHTsca and
the initial jet \pTisca for near (away) side jets.

\bigbreak

The angle \omegaHati is plotted as a function of \Dphii for different
values of \fii in the middle panel of Fig.~\ref{fig:f015}. The angle
\omegaHati is the same as \omegai for near side jets. For away side
jets, \omegaHati does not depend on \Dphii and its value is
$\arccot\fii$. The angle \omegaHati is the same as or wider than
\omegaTildei.

\subsubsection{Angle \texorpdfstring{\chii}{chii}}

The last variant \chii is a combination of \omegai and \bDphii.

\paragraph{\boldmath Ratio \kii.}
The ratio \kii is defined as
\begin{align}
\label{eq:ki}
\kii
\equiv \min(\fii, \gii)
= \begin{cases}
    \fii & \text{if $\Dphii \le \pi/2$}\\
    \gii & \text{otherwise},
\end{cases}
\end{align}
where \gii is defined in Eq.~\eqref{eq:gi}. The ratio \kii is the
ratio of the initial (minimizing) jet \pTisca and the initial \MHTsca
for near (away) side jets.

\paragraph{\boldmath Angle \chii.}
The angle \chii is defined such that $\tan\chii$ is the ratio of
$\sin\DphiTildei$ and \kii\footnote{As is the case for \bDphii, \chii
  is indeterminate if $\fii = 1$ and $\Dphii = \pi$ and can be defined
  as $\pi/2$.}:
\begin{align}
  \label{eq:tan_chi}
  \tan\chii \equiv \frac{\sin\DphiTildei}{\kii}.
\end{align}
The angle \chii ranges from 0 to $\pi/2$. The $\tan\chii$ is the ratio
of minimized \MHTsca and the initial (minimizing) jet \pTisca for near
(away) side jets.

\bigbreak

\begin{figure}[!h]
  \centering
    \includegraphics[scale=1.0]{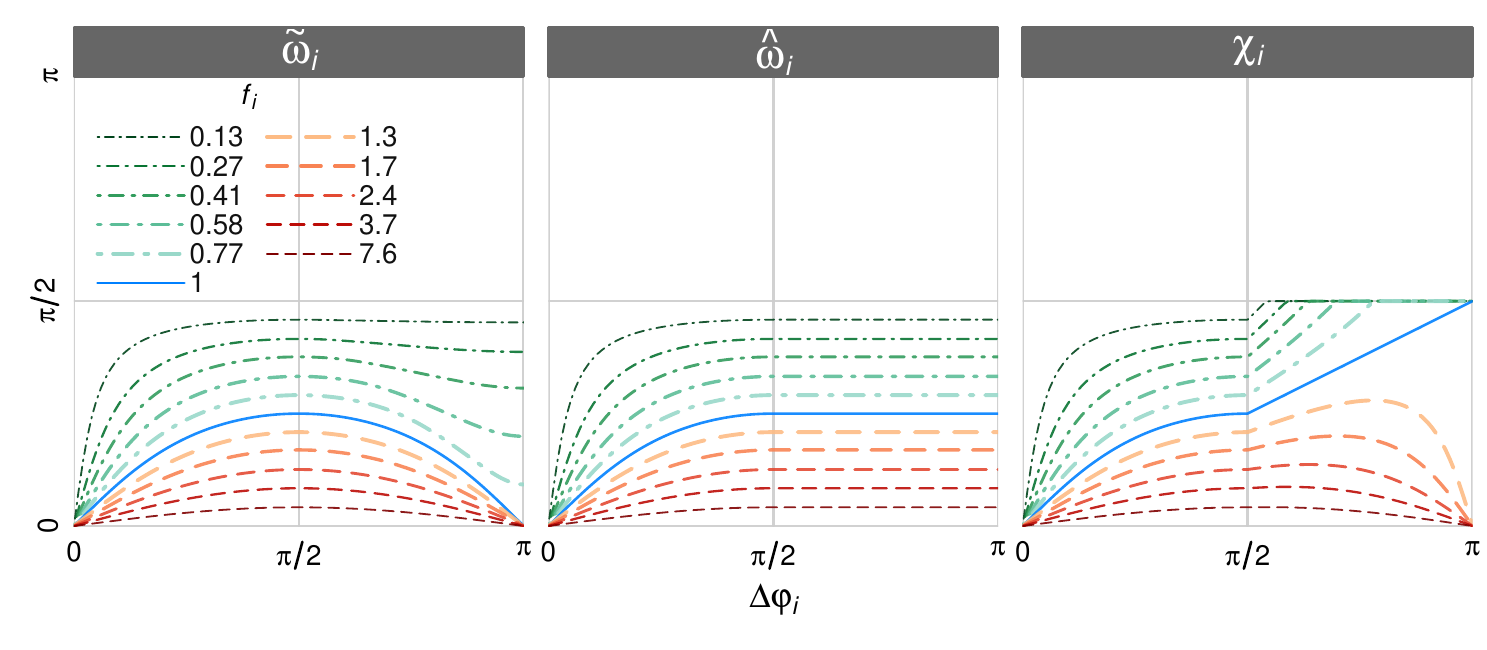}
    \caption{\label{fig:f015} The angles \omegaTildei, \omegaHati, and
      \chii as a function of \Dphii for different values of \fii.}
\end{figure}

In the rightmost panel of Fig.~\ref{fig:f015}, the angle \chii is
shown as a function of \Dphii for different values of \fii. It is not
smooth at $\Dphii = \pi/2$, but continuous. The angle \chii is the
same as \omegai for near side jets and is the same as \bDphii for away
side jets except \chii is capped at $\pi/2$. The angle \chii has the
advantage of \omegai for near side jets and of \bDphii for away side
jets.

\subsection{Alternative angular variables: \texorpdfstring{\minOmegaTilde}{minomega{\textasciitilde}}, \texorpdfstring{\minOmegaHat}{minomega{\textasciicircum}}, \texorpdfstring{\minChi}{minchi}}

Finally, we introduce the alternative angular variables. The three
variants of \omegai can be defined for each jet. The alternative
angular variables, to be defined for each event, are the minima of
these angles for all jets in the event:
\begin{align}
  \label{eq:min_omega_tilde}
  \minOmegaTilde &\equiv \min_{i\in\text{jets}}\omegaTildei \\
  \label{eq:min_omega_hat}
  \minOmegaHat &\equiv \min_{i\in\text{jets}}\omegaHati \\
  \label{eq:min_chi}
  \minChi &\equiv \min_{i\in\text{jets}}\chii.
\end{align}
The angles \omegaTildei, \omegaHati, and \chii are defined as their
tangents, respectively, in Eqs.~\eqref{eq:tan_omega_tilde},
\eqref{eq:tan_omega_hat}, and \eqref{eq:tan_chi}. There, the angles
are defined in terms of various scale factors and ratios defined
throughout the paper, such as $\sin\DphiTildei$, $\sin\DphiHati$,
\gii, and \kii.


Here, for convenience, we write down these angles in terms of only
\Dphii and \fii:
\begin{align}
  \omegaTildei
  =\arctan\left(
  \frac{\sqrt{1 + \left(\min\left(\fii,
      -\cos\Dphii\right)\right)^2+2\min\left(\fii,
      -\cos\Dphii\right)\cos\Dphii}}{\fii}
  \right),
\end{align}
\begin{align}
  \omegaHati
  =\arctan\left(
  \frac{\sin\left(\min(\Dphii, \pi/2)\right)}{\fii}
  \right),
\end{align}
\begin{align}
  \chii
  =\arctan\left(
  \frac{\sqrt{1 + \left(\min\left(\fii,
      -\cos\Dphii\right)\right)^2+2\min\left(\fii,
      -\cos\Dphii\right)\cos\Dphii}}{\min\left(\fii,
    \max\left(\fii+\cos\Dphii, 0\right)\right)}
  \right).
\end{align}
The numerators of all three fractions in the arguments of $\arctan$
range in $\left[0, 1\right]$. The denominators for \omegaTildei and
\omegaHati are positive, that is, $\fii > 0$. The denominator for
\chii is zero or positive. The angles \omegaTildei and \omegaHati
range in $\left[0, \pi/2\right)$. The angle \chii ranges in $\left[0,
    \pi/2\right]$.


\section{\boldmath Performance in simulated events}
\label{sec:roc_curves}

This section compares, in the simulated event samples described in
Section~\ref{sec:sample}, the performances of the three alternative
angular variables introduced in the previous section and two
conventional variables. The two conventional variables are \minbDphi,
which we reviewed in detail in Section~\ref{sec:min_bdphi_feature},
and \minDphiFour, which is defined as the minimum of the angles \Dphii
of up to four highest \pTsca jets in the event:
\begin{align}
  \label{eq:mindphi4}
  \minDphiFour \equiv \min_{i\in \{1,\cdots,\min(4, \njet)\}}\Dphii.
\end{align}
We picked \minDphiFour as the benchmark event variable of \Dphii for
the performance comparison. The \minDphiFour (with \Dphii defined with
\METvec) was used, for example, in Refs.~\cite{Sirunyan:2017kqq,
  Sirunyan:2017wif, Aaboud:2017hrg}.

The values of the five variables, \minDphiFour, \minbDphi,
\minOmegaTilde, \minOmegaHat, and \minChi, for the same event have the
following relations by definition:
\vspace*{-0.2cm} 
\begin{gather}
  \label{eq:minOmegaTilde_le_minOmegaHat}
  \minOmegaTilde \le \minOmegaHat \\
  \label{eq:minOmegaTilde_le_minChi}
  \minOmegaTilde \le \minChi \\
  \label{eq:minbDphi_le_minChi}
  \min(\minbDphi, \pi/2) \le \minChi\\
  \label{eq:minbDphi_le_minDphiFour}
  \minbDphi \le \minDphiFour.
\end{gather}

\vspace*{-0.6cm} 
\subsection*{\boldmath Event distributions}

\begin{figure}[!h]
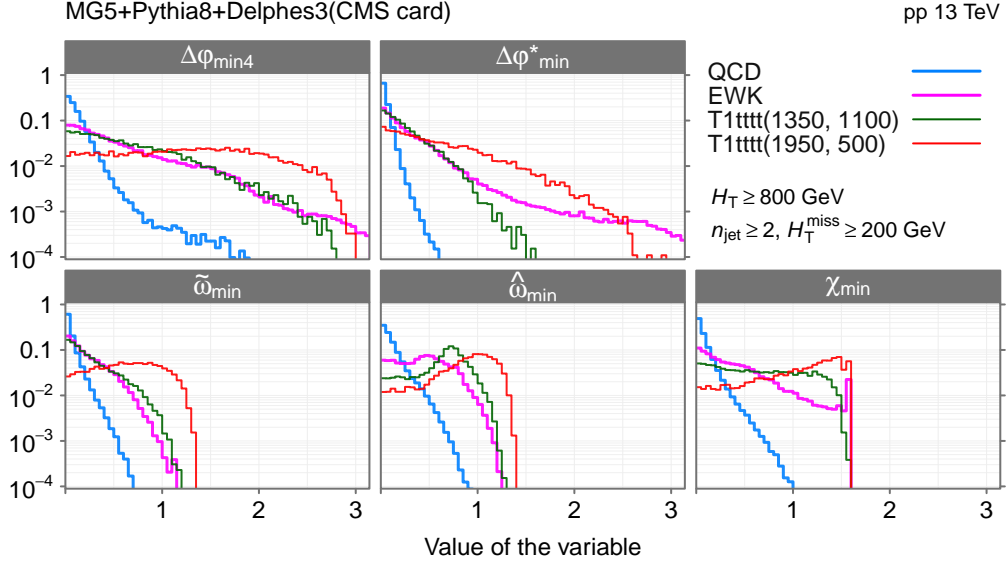

\centering
\includegraphics[scale=1.0]{{{g0510.xyplot.log10.val.var_htbin.var.htbin_800.process_010.delphes_cms}}}
\caption{\label{fig:g0510.delphes_cms} The distributions of QCD, EWK,
  T1tttt(1950,~500), and T1tttt(1350,~1100) events in the conventional
  variables (\minDphiFour, \minbDphi) and the alternative variables
  (\minOmegaTilde, \minOmegaHat, and \minChi). Each distribution is
  normalized to unity.}
\end{figure}

Figure~\ref{fig:g0510.delphes_cms} shows the event distributions of
the background and signal processes in the five variables. Each
distribution is normalized to unity. By definition, \minChi,
\minOmegaTilde, and \minOmegaHat range from zero to $\pi/2$, whereas
\minDphiFour and \minbDphi range from zero to $\pi$.

In all variables, the distributions of QCD multijet events have the
maximum at zero and quickly decrease roughly exponentially by orders
of magnitude, whereas the distributions of events for the other
processes are generally more spread. QCD multijet background events
can be, therefore, rejected by applying a threshold on one of these
variables.

The distribution of the QCD multijet events in \minChi decreases more
slowly than that in \minbDphi, which is unavoidable because of the
relation Eq.~\eqref{eq:minbDphi_le_minChi}. The advantage of \minChi
over \minbDphi is that, despite the high jet multiplicity, the
distribution of the signal events are roughly uniform, which can be
attributed to the improvement of \omegai for near side jets described
in Section~\ref{subsec:omegai}.

The distributions of EWK and signal events in \minOmegaHat have maxima
at larger values. These maxima appear because the value of \omegaHati
is close to $\arccot\fii$ except for narrow $\Dphii$ as we saw in
Section~\ref{subsubsec:omega_hati}. Because $\arccot\fii$ is a decreasing
function of \fii, the order of the processes for which these maxima
appear, i.e., EWK, T1tttt(1350,~1100), and T1tttt(1950,~500),
corresponds to the reverse order of the processes for which the maxima
of the distributions in \maxf appear
(Fig.~\ref{fig:g0110_520.n.process.maxF.delphes_cms}).

The distribution of the T1tttt(1950,~500) events in \minOmegaTilde has
a maximum at a larger value while the distributions of the
T1tttt(1350,~1100) and EWK events have a maximum at zero. This is
because the maximum of the \maxf distribution of the T1tttt(1950,~500)
events is less than unity but those for T1tttt(1350,~1100) and EWK are
greater than unity
(Fig.~\ref{fig:g0110_520.n.process.maxF.delphes_cms}).

The EWK events are distributed in the entire ranges of \minDphiFour,
\minbDphi, and \minbDphi, but not of \minOmegaTilde or \minOmegaHat.
In \minOmegaTilde and \minOmegaHat, the distributions of the EWK
events decrease somewhat more quickly than the distribution of the
T1tttt(1350,~1100) events and much more quickly than the distribution
of the T1tttt(1950,~500). EWK events can be, therefore, also reduced
by applying a threshold on \minOmegaTilde or \minOmegaHat.

\afterpage{\clearpage}

\subsection*{\boldmath ROC curves}

The ROC curves in Fig.~\ref{fig:g0420.roc.delphes_cms} demonstrate in
the simulated event samples that the alternative variables
\minOmegaHat and \minChi outperform the conventional variables
\minDphiFour and \minbDphi in rejecting QCD multijet background events
and that the alternative variables \minOmegaHat and \minOmegaTilde are
useful for reducing the total standard model background events.

\begin{figure}[!p]
\centering
  \begin{subfigure}[t]{1.0\textwidth}
    \centering
    \includegraphics[scale=1.0]{{{g0420.xyplot.log10qcd_eff.eff.roc_qcd_htbin.htbin.process2.var_110.delphes_cms}}}
    \caption{\label{fig:g0420.roc_qcd.delphes_cms}
      Against the selection efficiency of the QCD background events}
  \end{subfigure}
  \par\bigskip
  \par\bigskip
  \begin{subfigure}[t]{1.0\textwidth}
    \centering
    \includegraphics[scale=1.0]{{{g0420.xyplot.log10sm_eff.eff.roc_sm_htbin.htbin.process2.var_110.delphes_cms}}}
    \caption{\label{fig:g0420.roc_sm.delphes_cms} Against the
      selection efficiency of the standard model (QCD + EWK)
      background events}
  \end{subfigure}
  \caption{\label{fig:g0420.roc.delphes_cms} The ROC curves showing
    the selection efficiencies of T1tttt(1950,~500) and
    T1tttt(1350,~1100) signal events against the selection
    efficiencies of (a) the QCD multijet background events and (b) the
    total standard model (QCD + EWK) background events for the
    conventional variables (\minDphiFour, \minbDphi) and the
    alternative variables (\minOmegaTilde, \minOmegaHat, and \minChi)
    in three ranges of \HT. The markers indicate the values of the
    variables.}
\end{figure}

\afterpage{\clearpage}

Figure~\ref{fig:g0420.roc_qcd.delphes_cms} shows the ROC curves
against the selection efficiency of the QCD multijet background
events. The alternative variables \minOmegaHat and \minChi
considerably outperform the conventional variables for a wide range of
the efficiency for both signal models in all \HT ranges. For example,
at the same selection efficiency of the QCD multijet events with
$\minbDphi \ge 0.5$, \minChi can improve the signal efficiency for
T1tttt(1950,~500) by 9\% for the lowest \HT range, 18\% for the medium
\HT range, and 4\% for the highest \HT range and for
T1tttt(1350,~1100), 15\%, 20\%, and 12\% respectively in each \HT
range.

The alternative variable \minOmegaTilde performs, for
T1tttt(1950,~500), nearly as well as \minOmegaHat in the lowest \HT
range, but only as well as \minbDphi for the two highest \HT ranges.
For T1tttt(1350,~1100), \minOmegaTilde performs well only for the
tight background rejection in the lowest \HT range.

Figure~\ref{fig:g0420.roc_sm.delphes_cms} shows similar ROC curves but
against the selection efficiency of the total standard model
(QCD$+$EWK) background events. Here \minOmegaHat, overall, performs
the best among the five variables. At the same selection efficiency of
the background events with $\minbDphi \ge 0.5$, \minOmegaHat can
improve the signal efficiency for T1tttt(1950,~500) by 21\% for the
lowest \HT range, 25\% for the medium \HT range, and 18\% for the
highest \HT range and for T1tttt(1350,~1100), 24\%, 30\%, and 32\%
respectively in each \HT range. The variable \minOmegaTilde performs
nearly as well as \minOmegaHat for tight background rejection in all
\HT ranges for T1tttt(1950,~500). For T1tttt(1350,~1100),
\minOmegaTilde performs well only for the tight background rejection
in the lowest \HT range.

\section{\boldmath Summary}

We introduced three angular variables---\minOmegaTilde, \minOmegaHat,
and \minChi---as alternatives to \minbDphi as well as to \Dphii for
QCD multijet background event suppression in SUSY searches in
all-hadronic final states in proton-proton collisions at the LHC. We
demonstrated in simulated event samples that \minOmegaHat and \minChi
considerably outperform \minbDphi and \Dphii in rejecting QCD multijet
background events and that \minOmegaHat and \minOmegaTilde can also be
used to reduce the total standard model background events. In
particular, large improvements were observed for a signal model with
high jet multiplicity final states and with a compressed spectrum, in
which events tend to have low to medium \HT and low \MHTsca.

We evaluated the performance only for one detector model, specified in
the CMS detector configuration card in \DELPHES. In order to quickly
check the sensitivity of the performance to the detector model, we
repeated the performance evaluation with the ATLAS detector
configuration card, and we obtained nearly identical results, which is
shown in Appendix~\ref{appendix:atlas}. However, as discussed in
Appendix~\ref{appendix:jer}, the jet \pTsca resolutions of the real
CMS and ATLAS detectors might be better and have smaller tails than
those of both \DELPHES detector models, which potentially have a
certain impact on the performance.

Furthermore, we evaluated the performance only for two signal models
from a particular class of simplified SUSY models. These models have
many jets in the final states. As the variables are designed to
improve the signal acceptances of SUSY models with high jet
multiplicity, we can naively expect that \minOmegaHat and \minChi
perform well for other signal models with high jet multiplicity.
Nevertheless, it would be interesting to evaluate performances in a
wide range of signal models.

We aimed to develop variables for QCD multijet background event
suppression. However, \minOmegaHat and \minOmegaTilde turned out to be
also useful in reducing the EWK background events, including \znnjets
events, which are usually considered irreducible. It might be possible
to develop even better variables for EWK background event suppression
if dedicated development is carried out.

In the event samples, 23 pileup interactions on average were included
in each event. We have not studied the impact of varying the pileup
interactions, which might influence the results.

We focused on all-hadronic final states in this paper. However, in
searches in other final states, such as one lepton + jets + large
\METsca, it is also important to reject events with large \METsca
caused by a jet mismeasurement or neutrinos in hadron decays. In this
case, as mentioned in Section~\ref{sec:bdphi}, equivalent variables
can be defined by replacing \MHTvec with \METvec in the definitions of
\Dphii and \fii. It is also possible to add \pTvec of other objects,
e.g., the lepton \pTvec, to \MHTvec.

The variables \minOmegaTilde, \minOmegaHat, and \minChi are functions
of \Dphii and \fii of all jets in the event. In this paper, we
considered only a specific form of the functions, i.e., these
variables are each the minimum of some angles for all jets in the
event whose tangents are the ratios of variations of \MHTsca and jet
\pTsca, which can be written as functions of \Dphii and \fii. It is
probable that the best possible variable as a function of \Dphii and
\fii does not have this form. A variable with a somewhat different
form is described in Appendix~\ref{appendix:xi}. It might be possible
to develop better dimensionless variables as a function of \Dphii and
\fii, for example, with machine learning algorithms.

Increasing the signal-to-background ratio is one of the most
fundamental ways to improve the search for rare events. The
alternative variables introduced in this paper can increase the
signal-to-background ratio in SUSY searches. At the same time, it is
important to continue to develop new variables so as to make the best
use of the data at the LHC and future collider experiments.

\acknowledgments

All simulated events used in this paper were produced in Data
Intensive Computing Environment (DICE) at the University of Bristol
Particle Physics Group. This work was supported by Science and
Technology Facilities Council (UK).

\paragraph{Note.}
The code to calculate the alternative variables introduced in this
paper can be found at \url{https://github.com/TaiSakuma/altdphi}.


\appendix

\section{\boldmath Event selection with \texorpdfstring{\mmMHTsca}{mmMHT} and \texorpdfstring{$X_\text{min}$}{Xmin}}
\label{appendix:mmmht_event_selection}

It is suggested in Section~\ref{sec:min_minimized_mht} that \mmMHTsca,
defined in Eq.~\eqref{eq:minimum_minimized_mht}, can possibly replace
\MHTsca (or \METsca) in the event selection. We pursue this
possibility in this appendix. In addition to \mmMHTsca, we consider
another variable with the dimension of the momentum:
\begin{align}
  X_\text{min} \equiv \MHTsca\min_{i\in\text{jets}}(\tan\chii),
\end{align}
where \chii is defined in Eq.~\eqref{eq:tan_chi}.

\begin{figure}[!b]
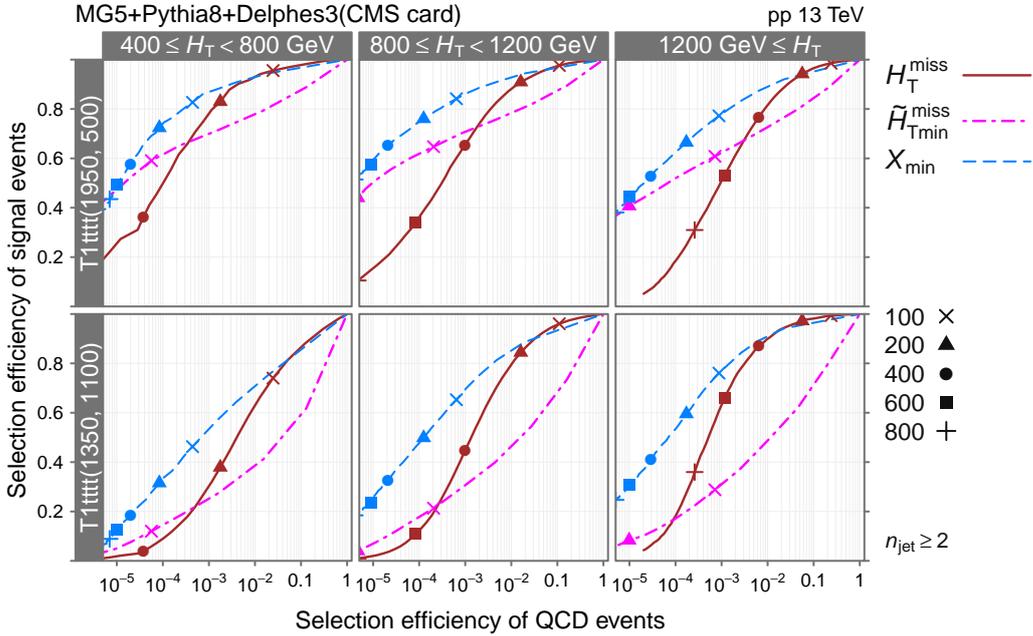

  \centering
  \vspace*{1mm}
\includegraphics[scale=1.0]{{{g0460.xyplot.log10eff.eff.roc_qcd_htbin.htbin.process2.var_043.delphes_cms}}}
\caption{\label{fig:g0460.delphes_cms} The ROC curves showing the
  selection efficiency of T1tttt(1950,~500) and T1tttt(1350,~1100)
  signal events against the QCD multijet background events for
  \MHTsca, \mmMHTsca, and $X_\text{min}$ in three ranges of \HT. The
  markers indicate the values of the variables in GeV.}
\end{figure}

Figure~\ref{fig:g0460.delphes_cms} compares the ROC curves for
\MHTsca, \mmMHTsca, and $X_\text{min}$. The event samples are the same
as the one described in Section~\ref{sec:sample} except the \MHTsca
requirement ($\MHTsca \ge 200$~GeV) is not applied. While \mmMHTsca
outperforms \MHTsca only for the tight selection, $X_\text{min}$
considerably outperforms \MHTsca and \mmMHTsca for a wide range of the
selection efficiency.

For example, in the bottom left panel of
Fig.~\ref{fig:g0460.delphes_cms}, the \MHTsca requirement ($\MHTsca
\ge 200$~GeV) reduces the signal events to 40\%. The variable
$X_\text{min}$ can keep them around 60\% at the same QCD background
efficiency---20\% improvement. This panel is for the lowest \HT range
for T1tttt(1350,~1100), the model with the compressed spectrum. Most
events in this model reside in this \HT range. It is generally
challenging to keep the signal acceptance of a compressed model large.
The variable $X_\text{min}$ or similar variables with the dimension of
the momentum can be useful alternatives to \MHTsca or \METsca for
searches for signals of models with compressed spectra.

\section{\boldmath Performance in Delphes samples with ATLAS card}
\label{appendix:atlas}

In the main body of this paper, we used the event samples simulated by
\DELPHES with the CMS detector configuration card as described in
Section~\ref{sec:sample}. We repeated the performance evaluation with
the ATLAS detector configuration card. We used the configuration file
\texttt{delphes\_card\_ATLAS\_PileUp.tcl} included in the \DELPHES
package with a slight modification. The same number of the pileup
interactions (23) were on average placed. The jet transverse momenta
\pTsca were corrected in the same way.

\begin{figure}[!h]
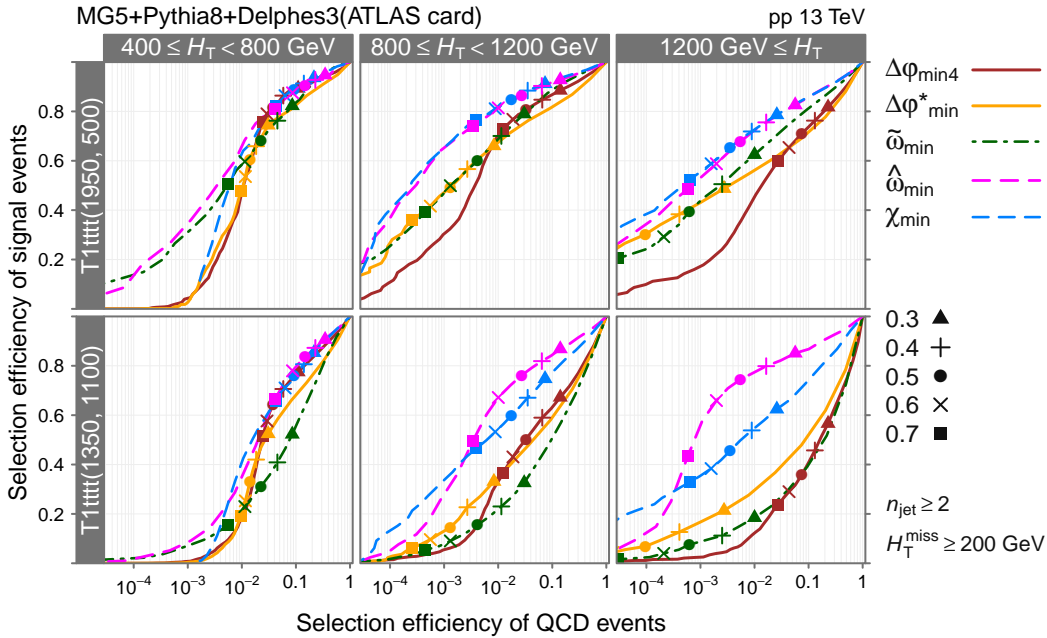

  \centering
  \vspace*{1mm}
  \includegraphics[scale=1.0]{{{g0420.xyplot.log10qcd_eff.eff.roc_qcd_htbin.htbin.process2.var_110.delphes_atlas}}}
  \caption{\label{fig:g0420.roc_qcd.delphes_atlas} The ROC curves for
    the five variables against the selection efficiency of the QCD
    multijet background events, the equivalent of
    Fig.~\ref{fig:g0420.roc_qcd.delphes_cms} but for the event samples
    simulated by \DELPHES with the ATLAS detector configuration card.}
\end{figure}

We obtained nearly identical results. For example,
Fig.~\ref{fig:g0420.roc_qcd.delphes_atlas} shows the ROC curves
equivalent to Fig.~\ref{fig:g0420.roc_qcd.delphes_cms} but with the
ATLAS detector configuration card. The two figures look alike. We can
draw the same conclusion as in Section~\ref{sec:roc_curves}.

However, in detail, all ROC curves in
Fig.~\ref{fig:g0420.roc_qcd.delphes_atlas} are somewhat shifted
towards right compared to those in
Fig.~\ref{fig:g0420.roc_qcd.delphes_cms}, indicating that the signal
efficiencies with the ATLAS card are somewhat lower than with the CMS
card at the same QCD background efficiency. This difference is in part
due to the difference in the jet \pTsca resolutions between the two
detector models, which will be discussed in
Appendix~\ref{appendix:jer}.

\section{\boldmath Jet \texorpdfstring{\pTsca}{pT} resolution}
\label{appendix:jer}

\begin{figure}[!t]
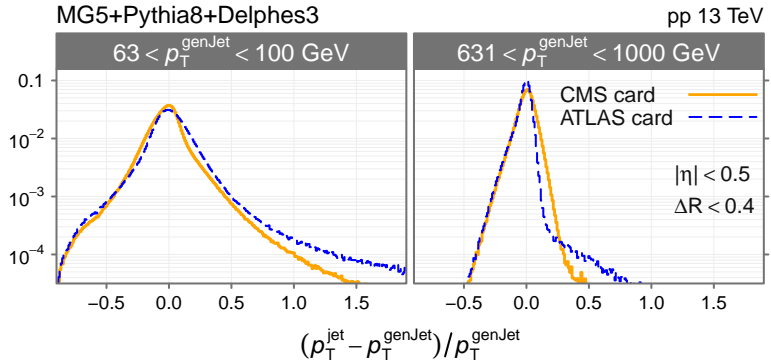

\centering
\vspace*{1mm}
\includegraphics[scale=1.0]{{{f110.xyplot.corr_norm_jet.jet_pt_63.card_010}}}
\caption{\label{fig:f110.jer} The jet \pTsca resolution: the
  distributions of ((jet \pTsca) - (generated jet \pTsca))/(generated
  jet \pTsca) in QCD multijet events simulated by \DELPHES with the
  CMS and ATLAS detector configuration cards in two ranges of
  generated jet \pTsca. A jet and generated jet within $\Delta R =
  \sqrt{\Delta\varphi^2 + \Delta\eta^2} < 0.4$ are matched. Each
  distribution is normalized to~unity.}
\end{figure}

Figure~\ref{fig:f110.jer} compares the jet \pTsca resolutions of the
two detector models, specified in the CMS and ATLAS configuration
cards in \DELPHES, in two ranges of the generated jet \pTsca. The
peaks are at zero because that is how we corrected jet \pTsca. The
distribution for the ATLAS configuration has a longer tail in the
right side, which caused the minor difference in the ROC curves
mentioned in Appendix~\ref{appendix:atlas}.

Somewhat equivalent figures in the CMS experiment can be found in
Figure~35 of Ref.~\cite{Khachatryan:2016kdb}, in which the \pTsca
ranges and the number of the pileup interactions are moderately
different from what we used in Fig.~\ref{fig:f110.jer}. However,
according to Figure~36 of the same reference, we can assume that the
impact of these differences is not significantly large. The collision
energy is also different. However, we can assume its impact is not
significantly large either.

On these assumptions, it can be inferred that the jet \pTsca
resolutions in the \DELPHES samples used in this paper are worse than
those in the CMS experiment. Furthermore, the shapes of the
distributions and the sizes of the tails are also different. In fact,
the difference between the distributions for the two detector
configurations of \DELPHES shown in Fig.~\ref{fig:f110.jer} is smaller
than the difference from the distributions in the CMS experiment shown
in Figure~35 of Ref.~\cite{Khachatryan:2016kdb}. These differences can
potentially have a certain impact on the performance of the
alternative variables introduced in this paper.

\section{\boldmath Alternative angular variable \texorpdfstring{$\xi$}{xi}}
\label{appendix:xi}

This appendix introduces yet another angular variable, which we let
$\xi$ denote. The ratio \hii is defined as
\begin{align*}
\hii =
\begin{cases}
    \gii & \text{for $i$ that minimizes $\sin\DphiTildei$}\\
    \fii & \text{otherwise}.
\end{cases}
\end{align*}
The variable $\xi$ is defined such that
\begin{align*}
  \tan\xi \equiv \frac{\displaystyle\min_{i\in\text{jets}}\sin\DphiTildei}{\displaystyle\max_{i\in\text{jets}}\hii}.
\end{align*}
This ratio is the equivalent of $1/\maxf$ after \MHTsca is minimized
by the variation of the jet \pTsca that minimizes \MHTsca most. In the
\DELPHES samples used in this paper, the variable $\xi$ performs
better than \minbDphi and \minDphiFour, but not better than
\minOmegaHat or \minChi.


\bibliographystyle{JHEP}
\bibliography{main}

\end{document}